\documentclass[fleqn,10pt]{wlscirep}
\usepackage[utf8]{inputenc}
\usepackage[T1]{fontenc}
\usepackage{subcaption}
\usepackage{siunitx}
\usepackage{placeins}
\usepackage{chngcntr}
\usepackage[version=4]{mhchem}

\title{NIDN: Neural Inverse Design of Nanostructures}

\newcommand{\gomez}{G\'{o}mez}

\newcommand{\etal}{\textit{et al.}}
\newcommand{\todo}[1]{\iffalse #1 \fi}
\newcommand{\tio}{\ce{TiO2}}
\newcommand{\ger}{\ce{Ge}}
\newcommand{\tape}{\ce{Ta2O5}}

\newcommand{\customPlotWidth}{0.8}

\author[1,*]{Pablo \gomez}
\author[1]{H\aa vard Hem Toftevaag}
\author[1]{Torbj\o{}rn Bogen-Stor\o{}}
\author[1]{Derek Aranguren van Egmond}
\author[2]{Jos\'{e} M. Llorens}
\affil[1]{European Space Agency, Advanced Concepts Team, Noordwijk, 2201AZ, The Netherlands}
\affil[2]{Instituto de Micro y Nanotecnología, IMN-CNM, CSIC (CEI UAM+CSIC) Isaac Newton,
8, E-28760, Tres Cantos, Madrid, Spain}

\affil[*]{Corresponding author, pablo.gomez@esa.int}

\begin{abstract}
In the recent decade, computational tools have become central in material design, allowing rapid development cycles at reduced costs. Machine learning tools are especially on the rise in photonics. However, the inversion of the Maxwell equations needed for the design is particularly challenging from an optimization standpoint, requiring sophisticated software. \\
We present an innovative, open-source software tool called \textit{Neural Inverse Design of Nanostructures} (NIDN) that allows designing complex, stacked material nanostructures using a physics-based deep learning approach. Instead of a derivative-free or data-driven optimization or learning method, we perform a gradient-based neural network training where we  directly optimize the material and its structure based on its spectral characteristics. NIDN supports two different solvers, rigorous coupled-wave analysis and a finite-difference time-domain method. The utility and validity of NIDN are demonstrated on several synthetic examples as well as the design of a \SI{1550}{\nano\metre} filter and anti-reflection coating. Results match experimental baselines, other simulation tools, and the desired spectral characteristics. \\
Given its full modularity in regard to network architectures and Maxwell solvers as well as open-source, permissive availability, NIDN will be able to support computational material design processes in a broad range of applications.
\end{abstract}
\begin{document}

\flushbottom
\maketitle
\thispagestyle{empty}


\section*{Introduction}
Computational tools have been playing an increasingly central role in material design in recent decades. With lower costs and rapid development cycles they are an efficient starting point in the material design process that can later be complemented with experimental validation and refinement \cite{wei2019machine}. Especially machine learning techniques have recently seen widespread adoption in applications ranging from predicting material properties to designing nanostructures \cite{liu2017materials,jiang2021deep} and this work aims to advance and support these efforts. \par

Brute-force parameter sweeps are increasingly being replaced by more sophisticated optimization methods in the design of metamaterials and nanophotonic devices \cite{Liu_Meta_Trans_FEM_GAN_2018, Ma_Meta_Refl_FreqDomain_VAE_2019, Tahersima_NanoPh_DM_FDTD_Backprop_2019, Piggott_NanoPh_DM_FDFD_GradDescent_2017, Rodriguez_NanoPh_DM_FDFD_Backprop_2021, sajedian2018finding, sajedian2019optimisation}. 
And, instead of letting human intuition and analytically derived parameters determine the architecture, nanostructured devices can now be autonomously designed for high performance, made possible by the exhaustive exploration of the design space that such algorithms perform \cite{Piggott_NanoPh_DM_FDFD_GradDescent_2017}.
In this context, this process of determining the physical characteristics from certain properties \cite{Molesky_2018_Inv_des_Nat_Phot}, e.g., finding the geometry of a structure for a given optical response, is called inverse design. 
Recently, there have been a lot of works on inverse design for nanophotonics using different techniques, such as generative adversarial networks \cite{Liu_Meta_Trans_FEM_GAN_2018}, variational auto-encoders \cite{Ma_Meta_Refl_FreqDomain_VAE_2019} and other machine learning and optimization methods \cite{wei2019machine, liu2017materials, jiang2021deep}.

One area where this has received a lot of attention is the space sector, where reducing the size, weight, power, and cost of objects put into orbit is not only a high priority, but can in many cases also determine the feasibility of the project. 
One example are photon sails, which must have sail thicknesses on the order of only a few micrometers (the IKAROS solar sail was only \SI{7.5}{\micro\metre} thick \cite{MORI_IKAROS_Sail_2010}) in order to avoid an unsuitable weight. 
Recently, there has been research on using metasurfaces for solar sails, made with both conventional design \cite{Ullery_Meta_Sail_FDTD_2018} and inverse design \cite{Jin_Meta_Sail_RCWA_AutoGrad_2020} methods. These metasurfaces might offer, for instance, attitude control \cite{Ullery_Meta_Sail_FDTD_2018} and mass reduction \cite{Jin_Meta_Sail_RCWA_AutoGrad_2020} while maintaining a high efficiency.

\begin{figure}[!htbp]
\centering
\includegraphics[width=\linewidth]{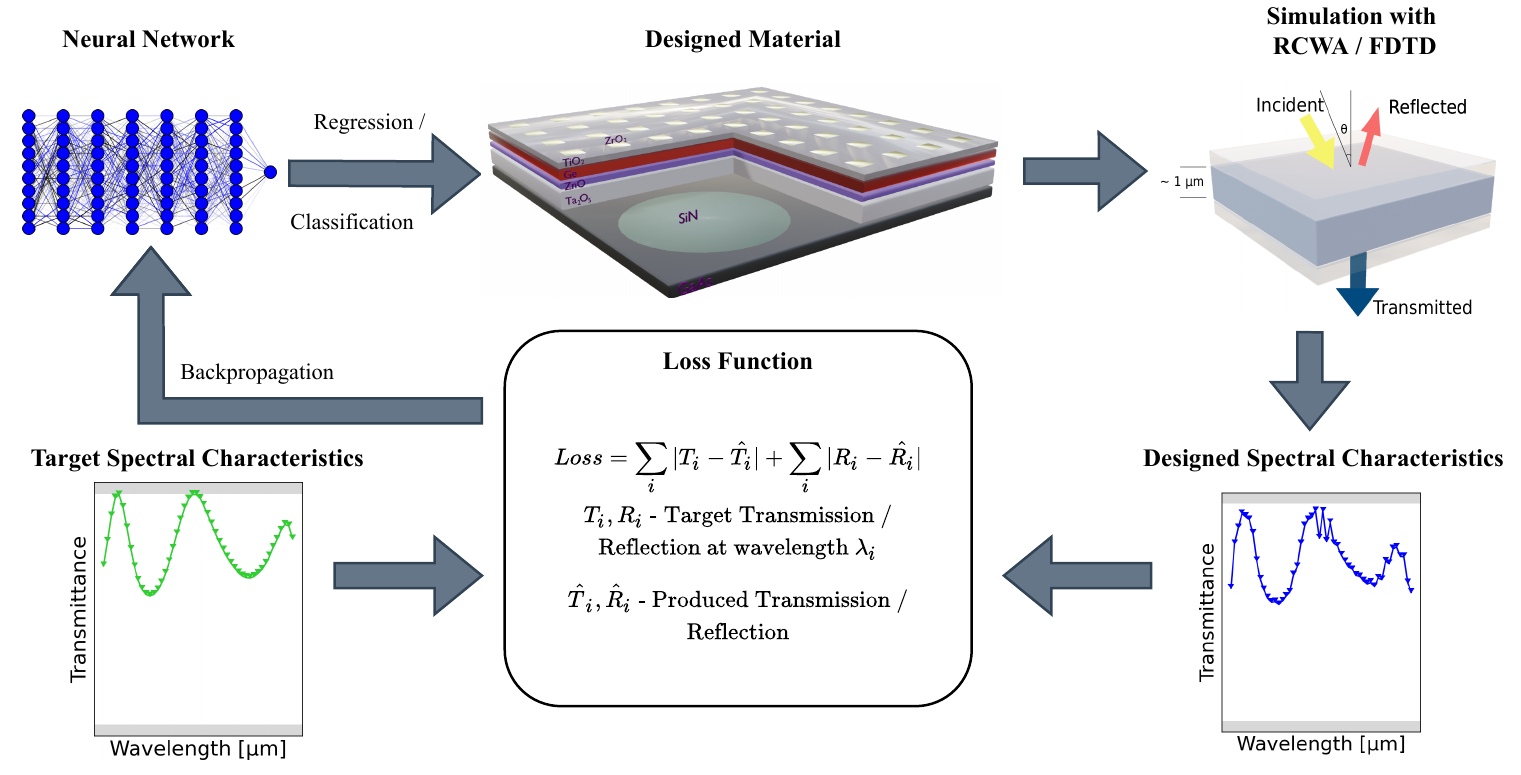}
\caption{Training setup for NIDN; regression and classification as well as RCWA and FDTD are described in detail in Methods. \label{fig:nidn}}
\end{figure}

Recent works have also applied metamaterial knowledge to the optical solar reflector (OSR), a structure placed on the external surface of a spacecraft that simultaneously reflects incoming sunlight and emit infrared radiation. 
Instead of using the conventional design of planar structures, Sun \textit{et al}.\ showed that a simple metasurface can reduce the thickness \cite{Sun_Meta_OSR_FEM-FDTD_2017, Sun_Meta_OSR_FEM-FDTD_2018}.
The terrestrial cousin of the OSR, radiative cooling structures, have followed a similar trend recently \cite{Hossain_Meta_OSR_FEM_2015}.

Overall, these methods rely on sophisticated software tools and datasets which has inspired efforts to make those available openly as, e.g., in the work of Jiang \etal{} \cite{jiang2020metanet}. In a similar vain, in this work, we present an open-source software package called \textit{Neural Inverse Design of Nanostructures} (NIDN) that allows designing nanostructures with neural networks using two different, fully differentiable implementations of Maxwell equations solvers based on the finite-difference time-domain (FDTD) method \cite{sullivan2013electromagnetic} and rigorous coupled-wave analysis (RCWA) \cite{moharam1981rigorous}. RCWA is a natural and efficient choice for stacked materials designed with NIDN. FDTD complements it by supporting design in the visible light range of the spectrum and - depending on geometry - providing higher accuracy \cite{han2014numerical}. Both enable the machine-learning-based design of multilayer materials with patterned nanoscale grids to achieve a desired spectrum of reflectance, transmittance and absorptance. Thus, NIDN has the potential to support a wide range of the aforementioned applications. In comparison to previous similar efforts such as those by Jiang \etal{} \cite{Jiang_Meta_OSR_FDTD_DBS_2021} we utilize a fundamentally different technique relying on directly encoding the material inside a neural network. \\
Building on advances in physics-based deep learning \cite{thuerey2021physics}, in NIDN, we directly optimize the nanostructure described by the neural network to achieve one specific spectrum. In this approach, no prior knowledge or training datasets are required as gradients are backpropagated through a (differentiable) numerical solver. Also, as the neural network does not need to solve the problem of fully describing the relationship between any material and spectral characteristics the task is more tractable. In many previous approaches \cite{Ma_Meta_Refl__VAE_2020,Nadell_Meta_Trans_CST_FFDS_2019}, neural networks effectively need to invert the Maxwell equations to match material and spectral characteristics to each other. In NIDN, by solving for a specific reflectance, transmittance and absorptance spectrum, we -- from a mathematical perspective -- constrain the inversion to a specific solution of the Maxwell equations. 
This furthermore avoids being limited to only a few degrees of freedom, such as placing a specific structure on top of another, in the material design \cite{Nadell_Meta_Trans_CST_FFDS_2019, Colburn_Meta_Refl_RCWA_Autodiff_2021}; instead we can fully explore the entire search space of structures.
Several nanosurface modification strategies have been employed to tune spectral response in optical devices. NIDN specifically excels in the parametric optimization of heterogeneous, multi-layer devices, which use local composition to tune electromagnetic properties. This is in contrast to the geometric structuring of a single material surface, which has limited local spectral tunability.

In fact, the neural network learn to produce a spatially continuous function describing the material permittivity. We demonstrate the validity and versatility of the approach by comparing to existing implementations that we built on. We showcase several examples ranging from simple uniform layers to practical applications related to photonic filter design and perfect anti-reflection. NIDN is also capable of running on graphics cards (GPUs) and is available open-source online.\footnote{\url{https://github.com/esa/NIDN} Accessed: 2022-04-21} The authors hope to build a community around the software, extending its capabilities and fidelity even further in the future.

\section*{Results}
In the following, we showcase results demonstrating NIDN's capabilities. In particular, we first validate the numerical Maxwell solvers in NIDN against baselines, then show the efficacy of the inverse design on some simple materials, and, finally, investigate two common design scenarios from the literature. 

\subsection*{Material Design Setup}
The following results were obtained using NIDN and its Maxwell solvers derived from the RCWA implementations by Jin \etal{} \cite{Jin_Meta_Sail_RCWA_AutoGrad_2020} and an adaption of an open-source implementation of FDTD in PyTorch 1.9.\footnote{\url{https://github.com/flaport/fdtd} Accessed: 2022-04-21} Both are described in more detail in the Section Methods. NIDN utilizes PyTorch 1.9 to enable automatic differentiation and neural network training to design materials of stacked layers such as shown in Figure \ref{fig:nidn}. Details on neural network architectures and training as well as the detailed training setup are given in the Section Methods. Data on material permittivity are obtained from an online database.\footnote{\url{https://refractiveindex.info} Accessed: 2022-04-21} More detailed info on specific references for each material are provided in the integrated database of materials in NIDN.\footnote{\url{https://github.com/esa/NIDN/tree/main/nidn/materials/data} Accessed 2022-04-21} \\
Permittivity between provided measurement points is interpolated linearly. Layer thickness if not specified otherwise is \SI{1}{\micro\metre}. Grid dimensions if not specified are \SI{1}{\micro\metre} $\times$ \SI{1}{\micro\metre}. In general, NIDN supports two modes of inverse design which are described in more detail in the Section Methods: The first one is referred to as regression, where the permittivity of the material is predicted within some value range to investigate hypothetical materials achieving the target spectral characteristics. If not specified otherwise, the range for relative permittivity $\epsilon_r$ considered by the network is $[0.01 , 20 + 3i]$. All references to permittivity refer to relative permittivity. The second approach is called classification and the permittivity in this case is chosen as a linear combination of reported permittivity of real materials. All considered materials are listed in the codebase. At present, only normal incidence is considered. Figure \ref{fig:nidn} gives a high-level overview of the training process. \\
The number of training iterations depended on the task and solver; for RCWA between 1000 and 5000 training iterations were used, for FDTD not more than 150 iterations due to the higher computation time. No dispersion model is used and frequencies are computed independently. For a single uniform layer, a single training iteration with FDTD requires about eight seconds per frequency, with RCWA it takes six milliseconds per frequency per iteration. L1 errors reported in the figures refer to the mean absolute error between the target and obtained spectrum. Inversion with FDTD used a coarser grid in the numerical discretization than for validation to keep the computational cost manageable (32 instead of 100).

\begin{figure}[!htbp]
\centering
\subcaptionbox{NIDN RCWA \tio{} Layer Spectrum}
{\includegraphics[width=\customPlotWidth\linewidth]{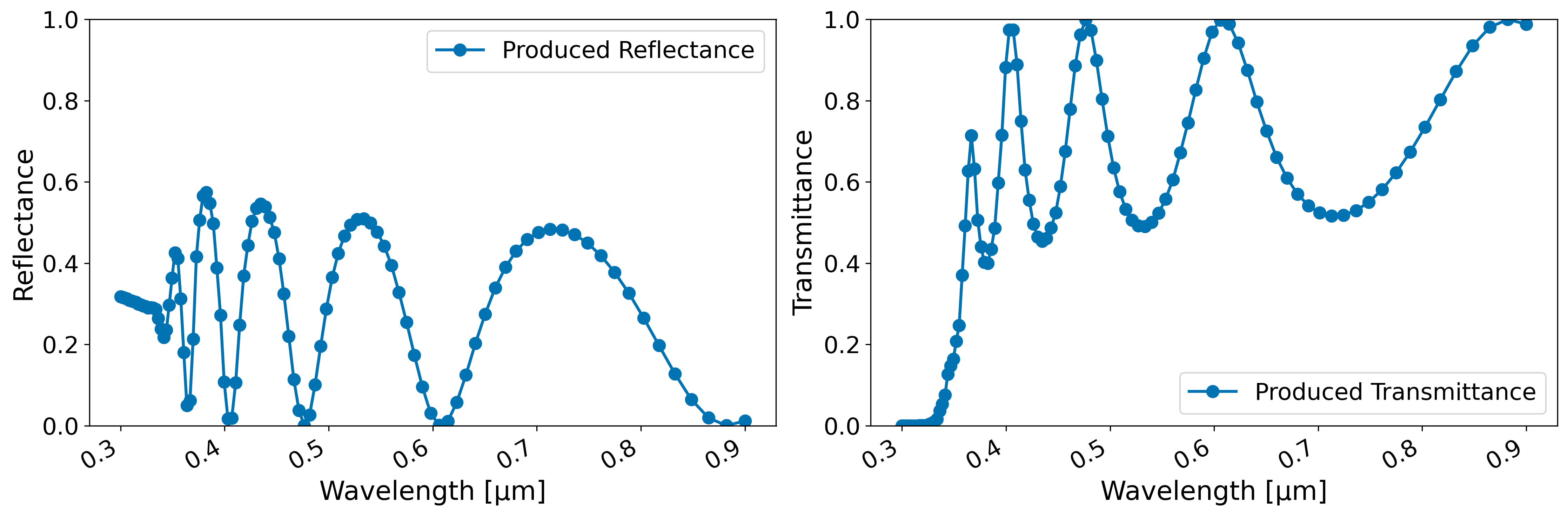}}
\subcaptionbox{NIDN FDTD \tio{} Layer Spectrum}
{\includegraphics[width=\customPlotWidth \linewidth]{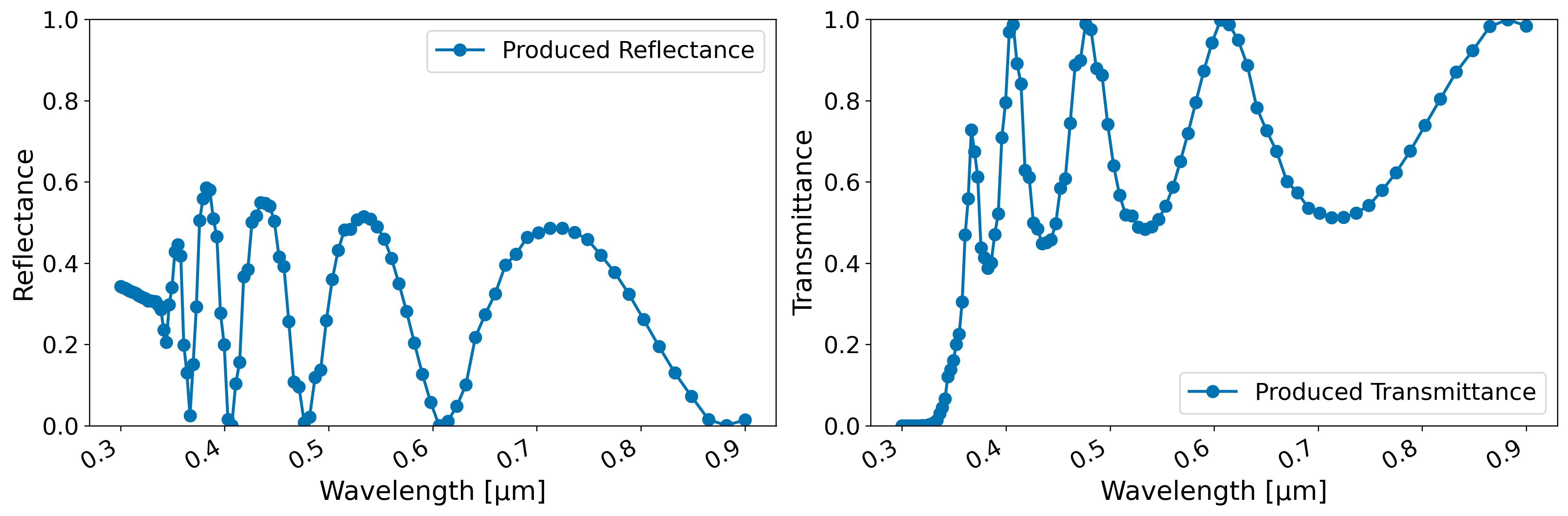}}
\subcaptionbox{Experimental results by Wojcieszak \etal{}\cite{Wojcieszak2016TransmissionTiO2} (reprinted with permission)}
{\includegraphics[width=0.4\linewidth]{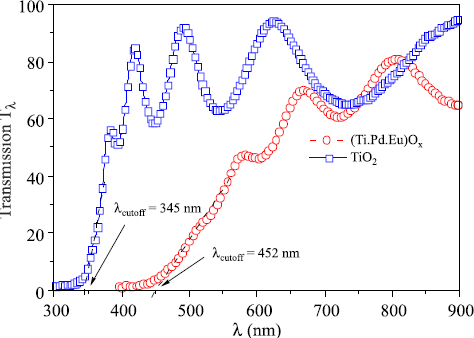}}
\caption{Comparison of the transmittance of a \SI{380}{\nano\metre} thick layer of \tio{} in NIDN and experimentally measured}
\label{fig:fdtd1}
\end{figure}

\begin{figure}[!htbp]
\centering
\subcaptionbox{Target and Obtained Spectrum}
{\includegraphics[width=\customPlotWidth\linewidth]{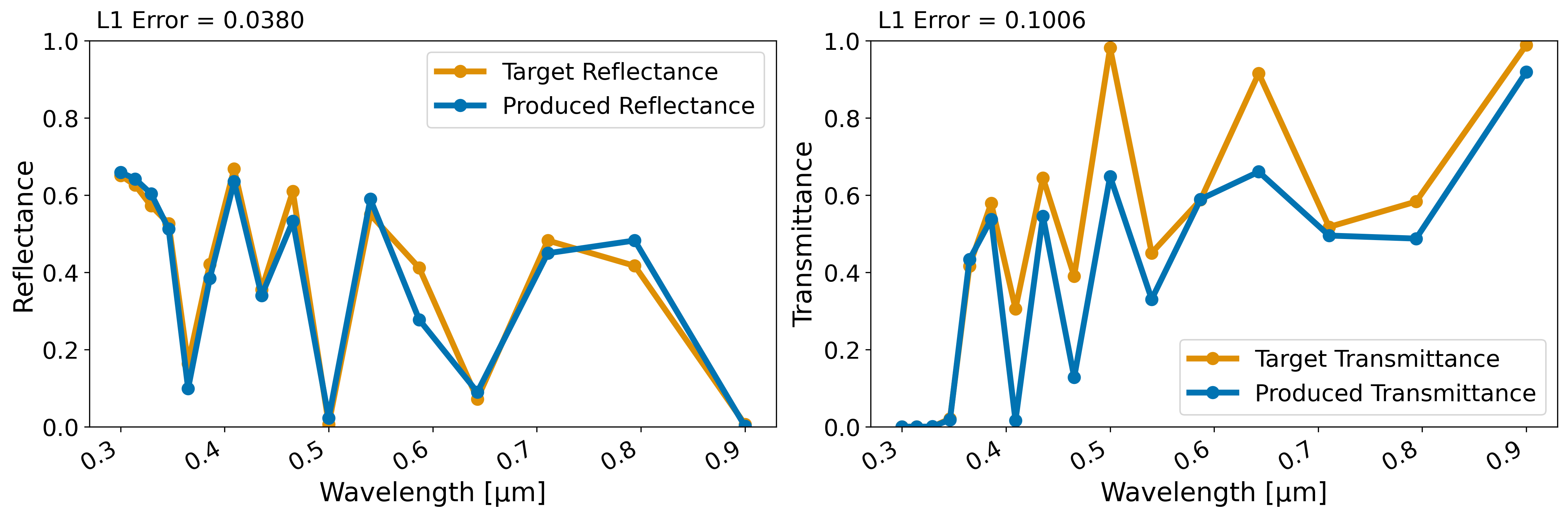}}
\subcaptionbox{Predicted Permittivity}
{\includegraphics[width=\customPlotWidth\linewidth]{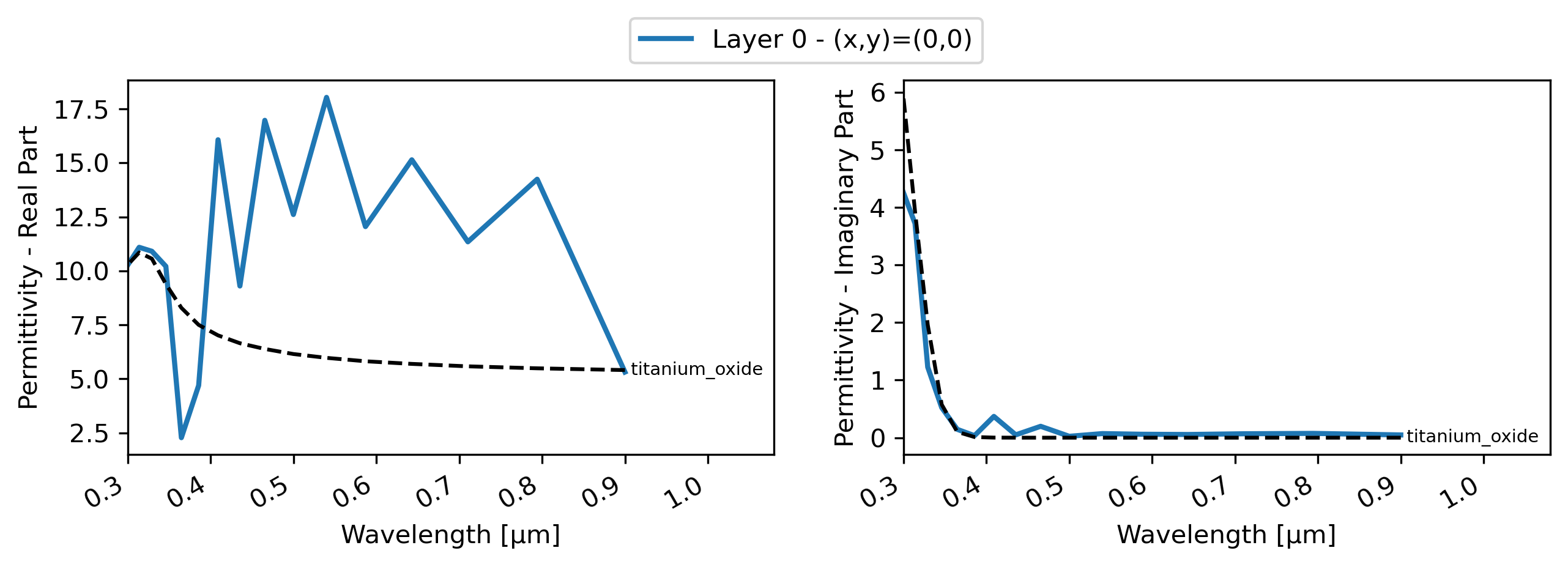}}
\caption{Inversion results for a single-layer \tio{} material using FDTD and regression; a) displays target and produced spectra, b) shows the utilized permittivity; differences to Figure \ref{fig:fdtd1} are due to smaller number of discretization grid points}
\label{fig:fdtd_reg1}
\end{figure}

\subsection*{Validation and Numerical Stability}
NIDN contains a fully differentiable RCWA implementation that is derived from the Python module GRCWA\cite{Jin_Meta_Sail_RCWA_AutoGrad_2020}. In contrast to GRCWA, NIDN utilizes PyTorch. Extensive tests are performed to validate both implementations against the original implementations and results of these can be found in the Supplementary Information. \\
NIDN adapts the FDTD implementation from the Python module with the same name (fdtd\footnote{\url{https://github.com/flaport/fdtd} Accessed: 2022-04-21}) but adds support for gradient flow through it and performs post processing. A comparison with an experimental baseline is given to validate the implemented post processing. Further, we directly compare results with the original Python module to validate our changes to it. \\
A comparison with an experimental measurement of the transmission spectrum of \tio{} is performed for both. The \tio{} layer in the experiment by Wojcieszak \etal{}\cite{Wojcieszak2016TransmissionTiO2} was \SI{380}{\nano\metre} thick and transmission of wavelengths between \SIlist{300;900}{\nano\metre} were measured using a spectrophotometer. Figure \ref{fig:fdtd1} displays transmittance obtained with NIDN RCWA and FDTD using continuous waves with wavelengths from \SIrange{300}{900}{\nano\metre} as well as the experimental result. Overall, it can be seen that the transmittance in NIDN resembles the one observed in the experiment. Point clustering in the FDTD version is a discretization effect due to the number of grid points used in the simulation.

\begin{figure}[!htbp]
\centering
\subcaptionbox{Target and Obtained Spectrum}
{\includegraphics[width=\customPlotWidth\linewidth]{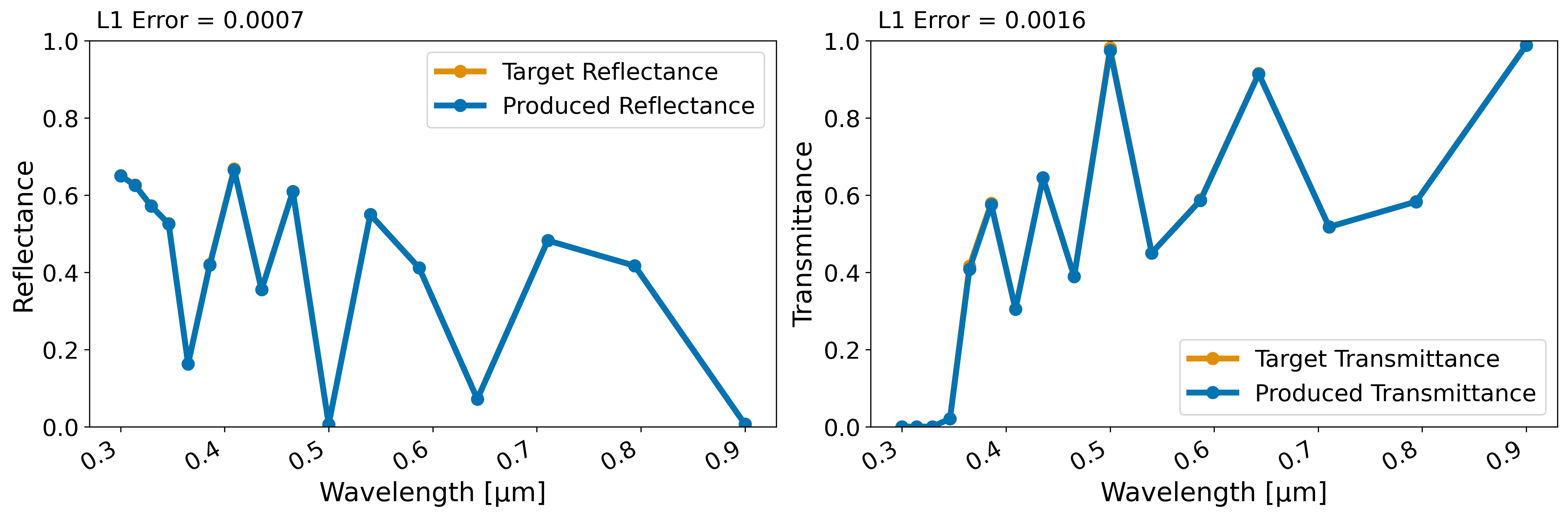}}
\subcaptionbox{Predicted Permittivity}
{\includegraphics[width=\customPlotWidth\linewidth]{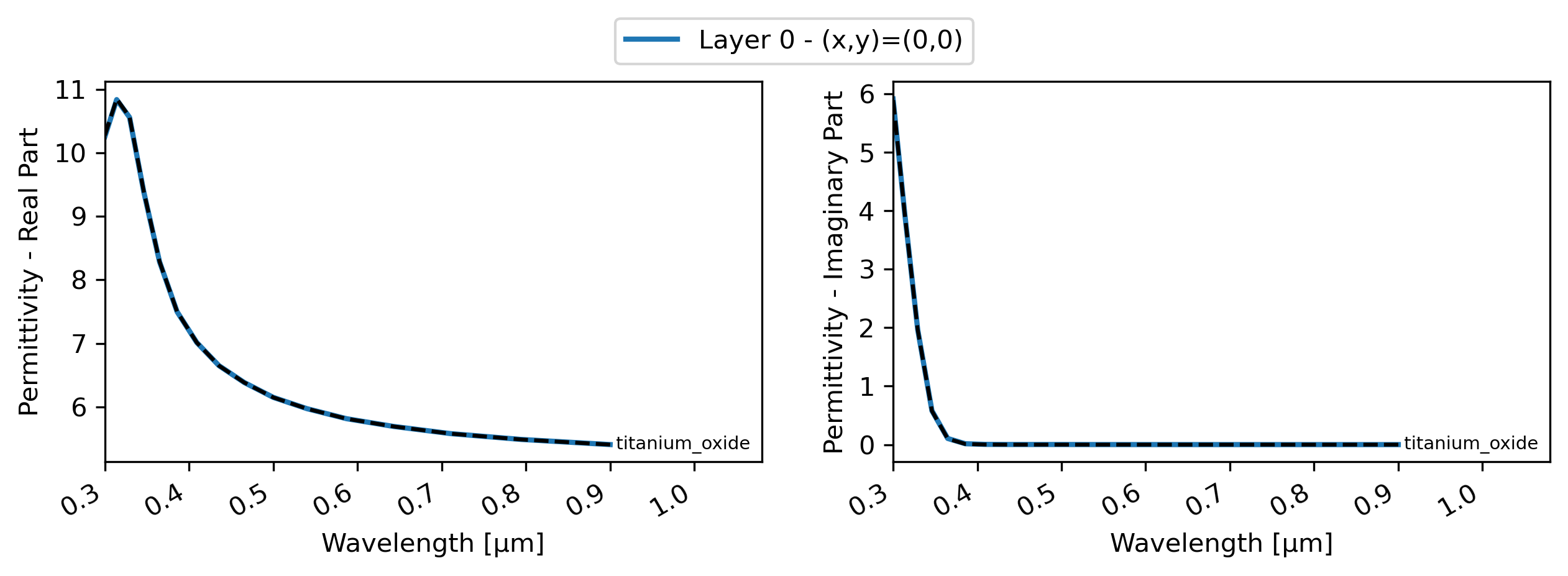}}
\caption{Inversion results for a single-layer \tio{} material using FDTD and classification; a) displays target and produced spectra, b) shows the utilized permittivity; differences to Figure \ref{fig:fdtd1} are due to smaller number of discretization grid points}
\label{fig:fdtd_class1}
\end{figure}

\subsection*{Material and Structure Inference}
As a first demonstration of NIDN's capabilities, we present results on some rudimentary examples. These examples aim to reconstruct a material for which the spectral characteristics were computed with RCWA / FDTD, thus the inversion is certain to have a solution obtainable by the network. In particular, these results show that, depending on the complexity of the spectral characteristics, almost perfect results can be obtained. Furthermore, we can investigate how unique solutions are by studying the similarity of designed materials with the ground-truth baseline.
\subsubsection*{Uniform \tio{} Layer Inversion}
First, we show that NIDN is capable of recreating the permittivity of a single layer of \tio, with FDTD in the visible range of the spectrum (for RCWA a more complex three-layer example is tested). \\
Figure \ref{fig:fdtd_reg1} showcases these results. Note that to save computational time, a limited number of frequency points, i.e. 20, were investigated. A clear convergence towards the ground truth is visible. However, the predicted permittivity clearly differs from the real one.

Even though the inversion with the regression approach was successful, it was not constrained to realistic materials. In Figure \ref{fig:fdtd_class1} we show results for the classification approach for this single-layer material. Overall, the inversion is successful and the obtained spectral characteristics are even closer to the target. Furthermore, as seen in Figure \ref{fig:fdtd_class1}b, the used epsilon value now corresponds well to the experimental permittivity of \tio{}.

\begin{figure}[!htbp]
\centering
\subcaptionbox{Target and Obtained Spectrum}
{\includegraphics[width=\customPlotWidth\linewidth]{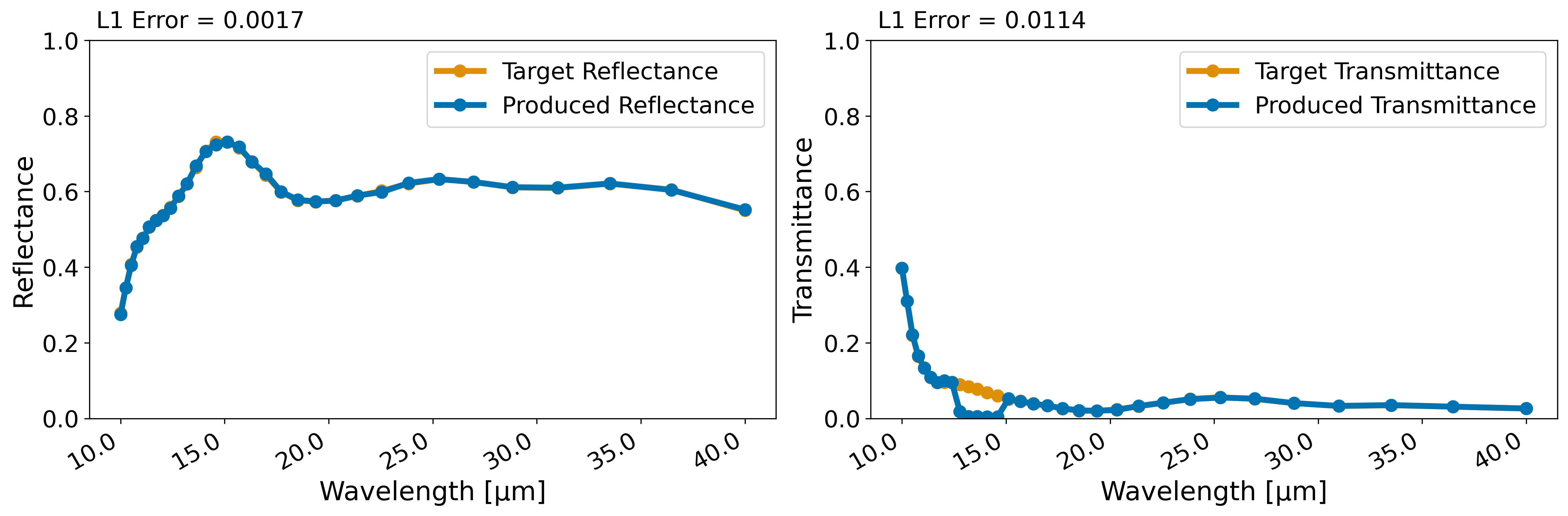}}
\subcaptionbox{Predicted Permittivity}
{\includegraphics[width=\customPlotWidth\linewidth]{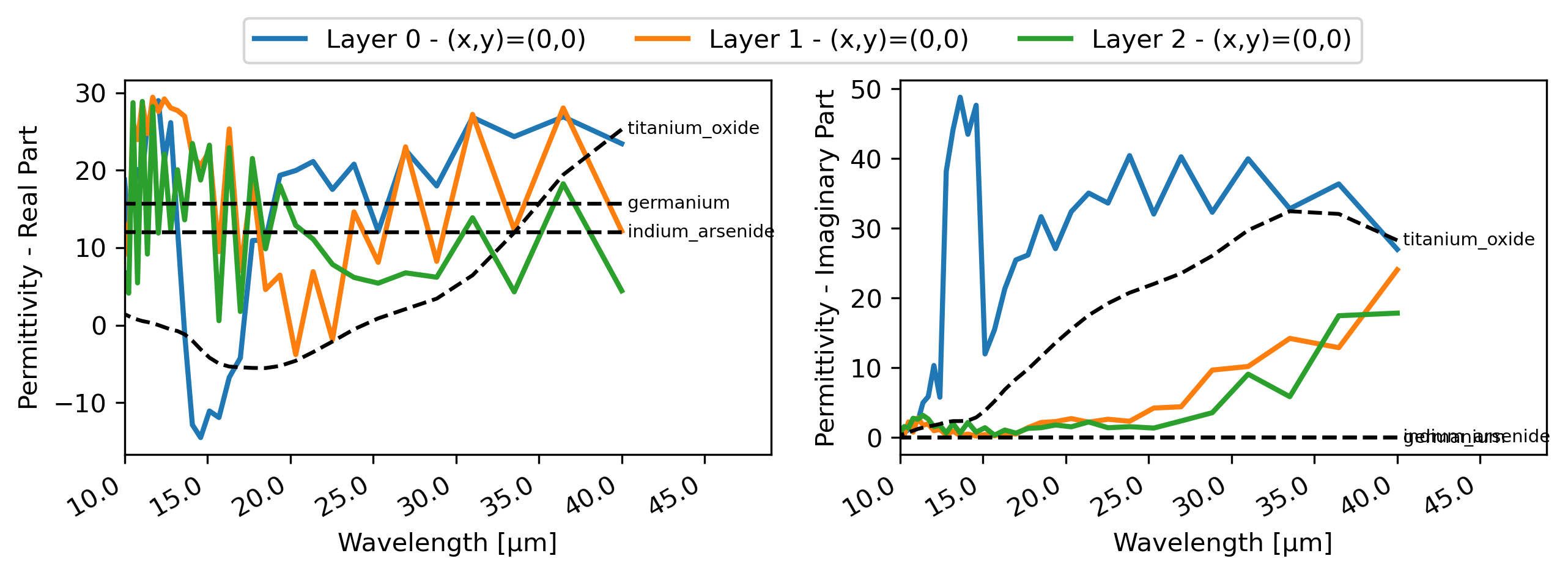}}
\caption{Inversion results for a three-layer material using RCWA and regression; a) displays target and produced spectra, b) shows the utilized permittivity for each layer}
\label{fig:reg2}
\end{figure}

\begin{figure}[!htbp]
\centering
\subcaptionbox{Target and Obtained Spectrum}
{\includegraphics[width=\customPlotWidth\linewidth]{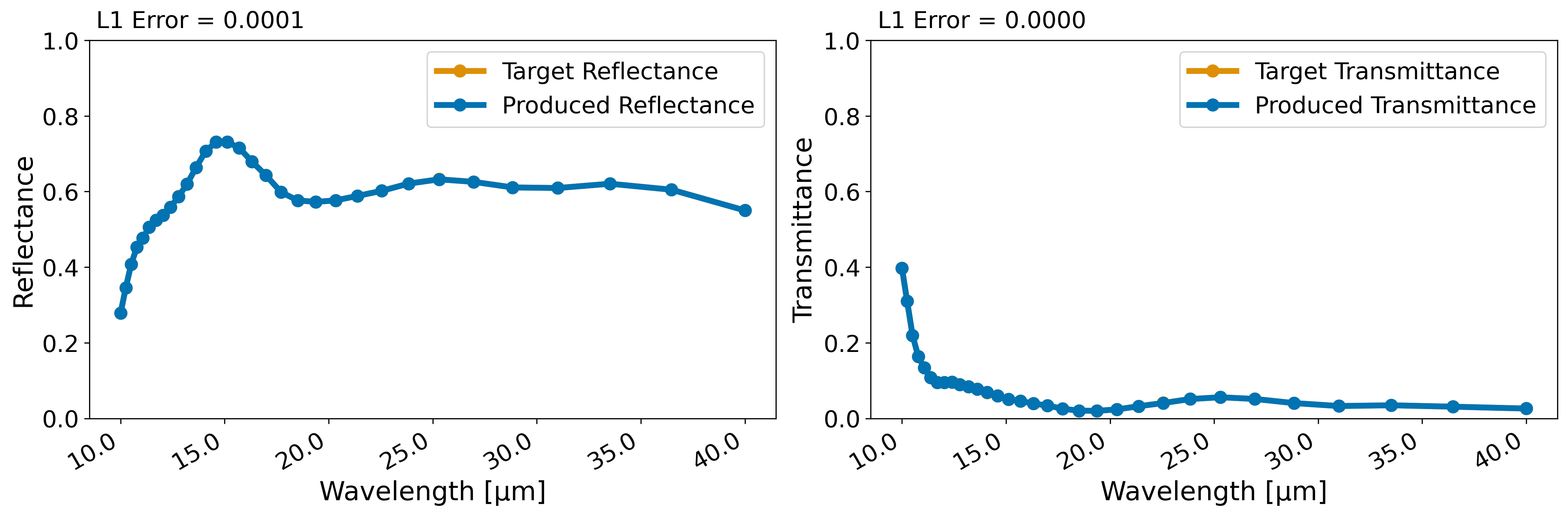}}
\subcaptionbox{Predicted Permittivity}
{\includegraphics[width=\customPlotWidth\linewidth]{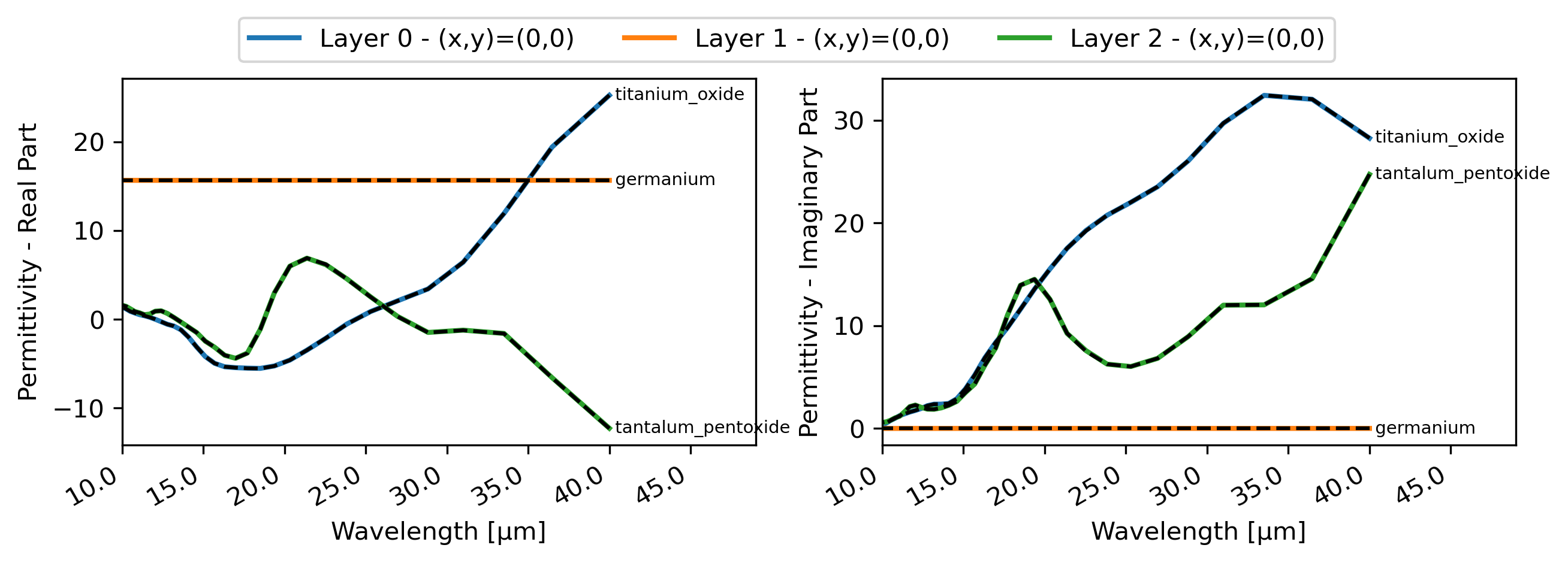}}
\caption{Inversion results for a three-layer material using RCWA and classification; a) displays target and produced spectra, b) shows the utilized permittivity for each layer}
\label{fig:class2}
\end{figure}

\subsubsection*{Three-Layer Uniform Material}
The second example is the design of a three-layer material consisting of \tio, Germanium (\ger) and \tape. Figure \ref{fig:reg2} displays the regression results with RCWA for this material. Again, the design of the spectral properties is successful. However, permittivities of the different layers are notably further from real materials. The obtained permittivity is not resemblant of the used materials. \\
In contrast, with the classification and RCWA approach, the restriction for the permittivity actually leads to the network learning the permittivity of the utilized materials for each layer, as can be seen in Figure \ref{fig:class2}. Furthermore, the achieved spectral characteristics also almost perfectly match the design goal of the ground truth. 

\begin{figure}[!htbp]
\centering
\subcaptionbox{Target and Obtained Spectrum}
{\includegraphics[width=\customPlotWidth\linewidth]{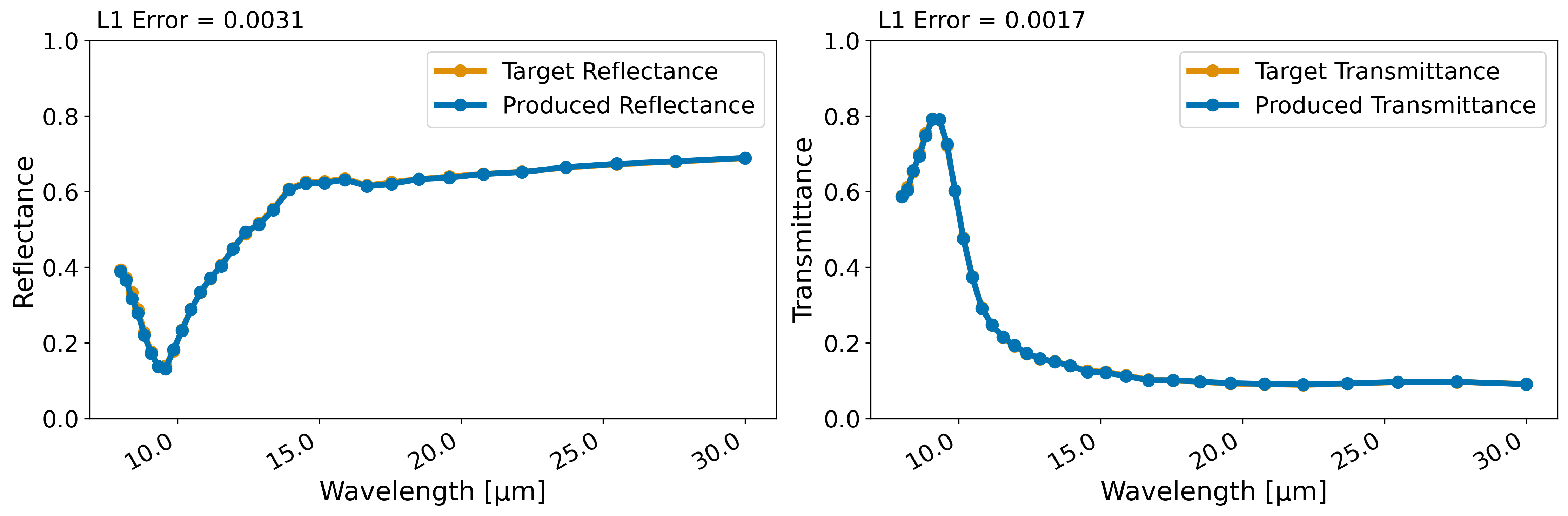}}
\subcaptionbox{Predicted Permittivity}
{\includegraphics[width=\customPlotWidth\linewidth]{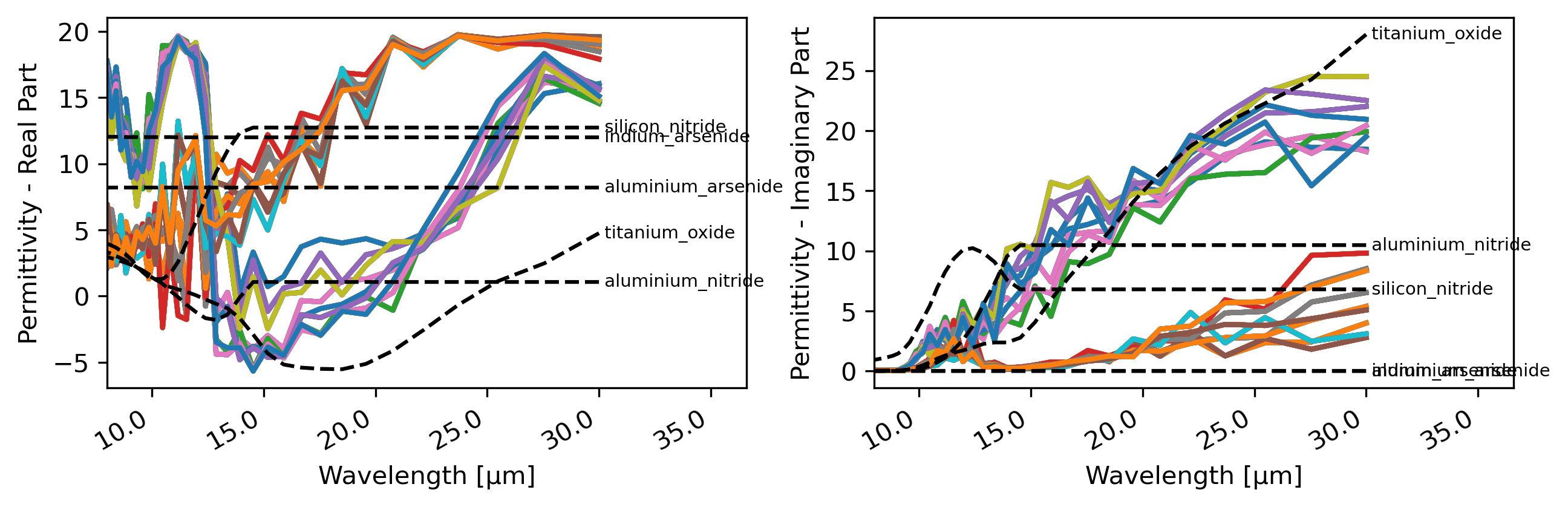}}
\caption{Inversion results for a two-layer patterned material using RCWA and regression; a) displays target and produced spectra, b) each line shows the utilized permittivity for a grid cell}
\label{fig:reg3}
\end{figure}
\begin{figure}[!htbp]
\centering
\subcaptionbox{Target and Obtained Spectrum}
{\includegraphics[width=\customPlotWidth\linewidth]{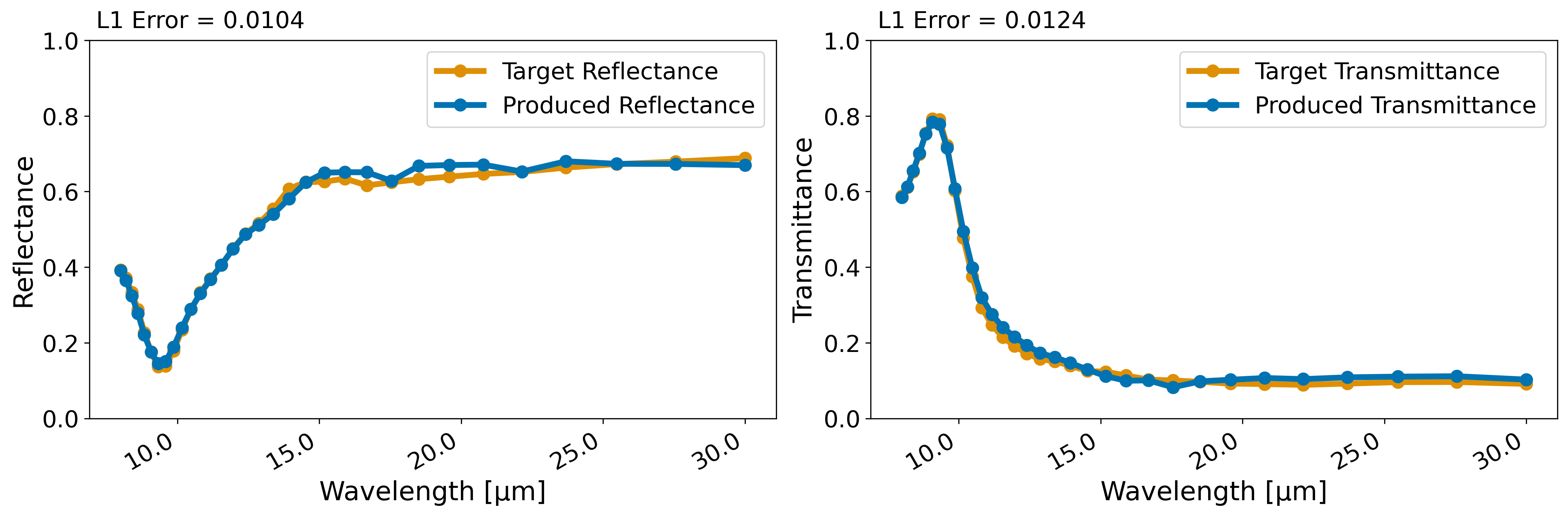}}
\subcaptionbox{Predicted Permittivity}
{\includegraphics[width=\customPlotWidth\linewidth]{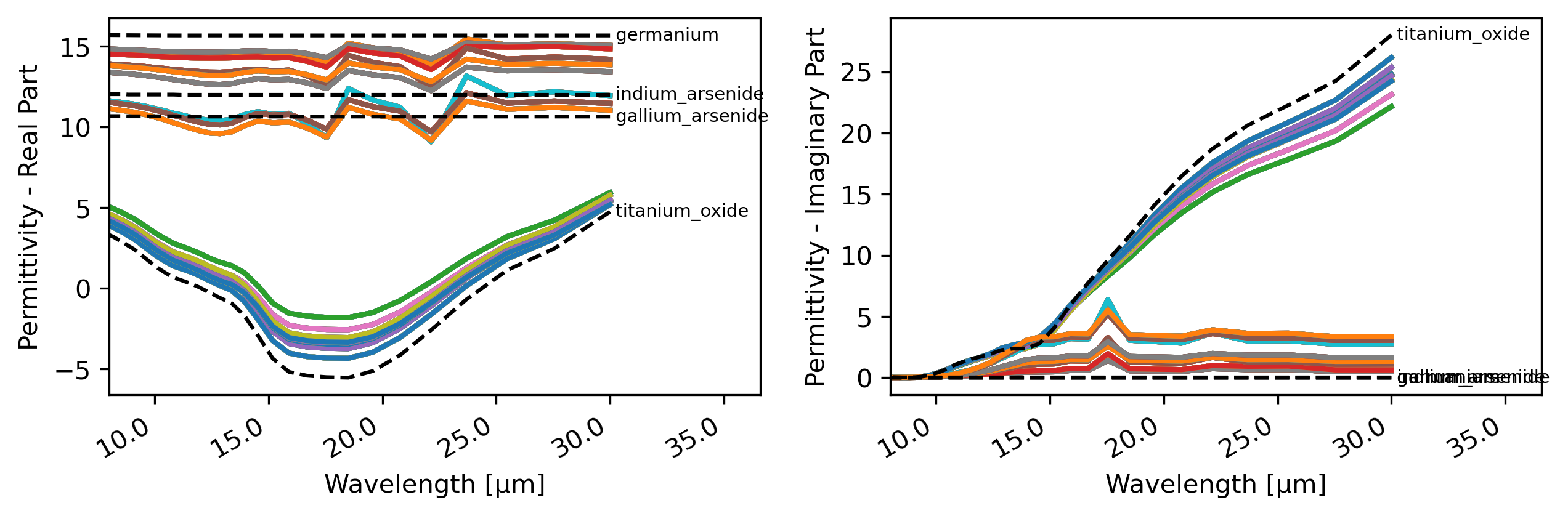}}
\caption{Inversion results for a two-layer patterned material using RCWA and classification; a) displays target and produced spectra, b) each line shows the utilized permittivity for a grid cell}
\label{fig:class3}
\end{figure}

\subsubsection*{Two-Layer Patterned Material}
The last case based on a ground truth from the forward models (i.e., the RCWA and FDTD implementations) is a two-layer patterned material consisting of \tio{} and \ger{} with a square pattern in the center similar to those shown in the Supplementary Information. In the upper layer the square is made of \tio{} and surroundings of \ger{} and vice versa for the bottom layer. This test case is only studied with RCWA as support for patterned layers is currently still in development for FDTD. \\
Figure \ref{fig:reg3} displays the obtained spectral characteristics with regression and RCWA. The desired spectrum is reproduced very well as can be seen in Figure \ref{fig:reg3}a. However, the utilized permittivity is not close to any material available in NIDN as  \ref{fig:reg3}b shows and the reconstructed pattern is not resemblant of the ground truth. \\
For the classification approach and RCWA, shown in Figure \ref{fig:class3}, comparable errors are obtained. In this case -- as can be seen in Figure \ref{fig:class3}b -- the permittivity chosen by the network is close to those of used in the ground truth, i.e., \tio{} and \ger{}. The pattern of the ground truth was not reconstructed implying that the network found a good solution without it.

\begin{figure}[!htbp]
\centering
\subcaptionbox{Target and Obtained Spectrum}
{\includegraphics[width=\customPlotWidth\linewidth]{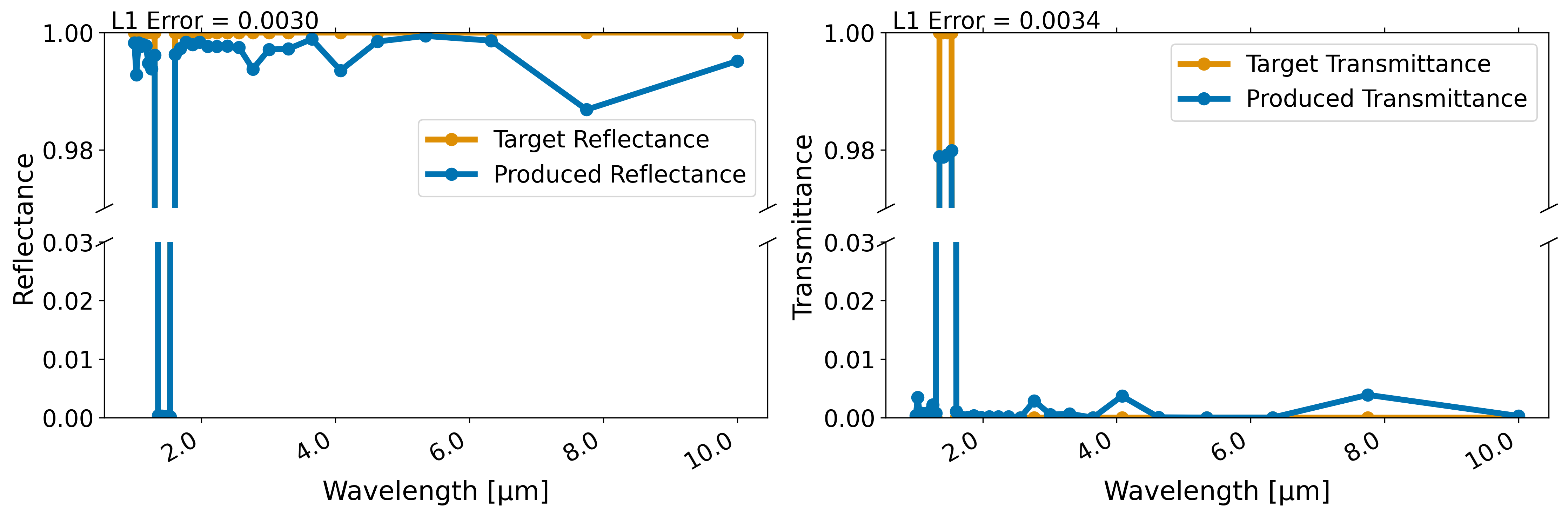}}
\subcaptionbox{Predicted Permittivity}
{\includegraphics[width=\customPlotWidth\linewidth]{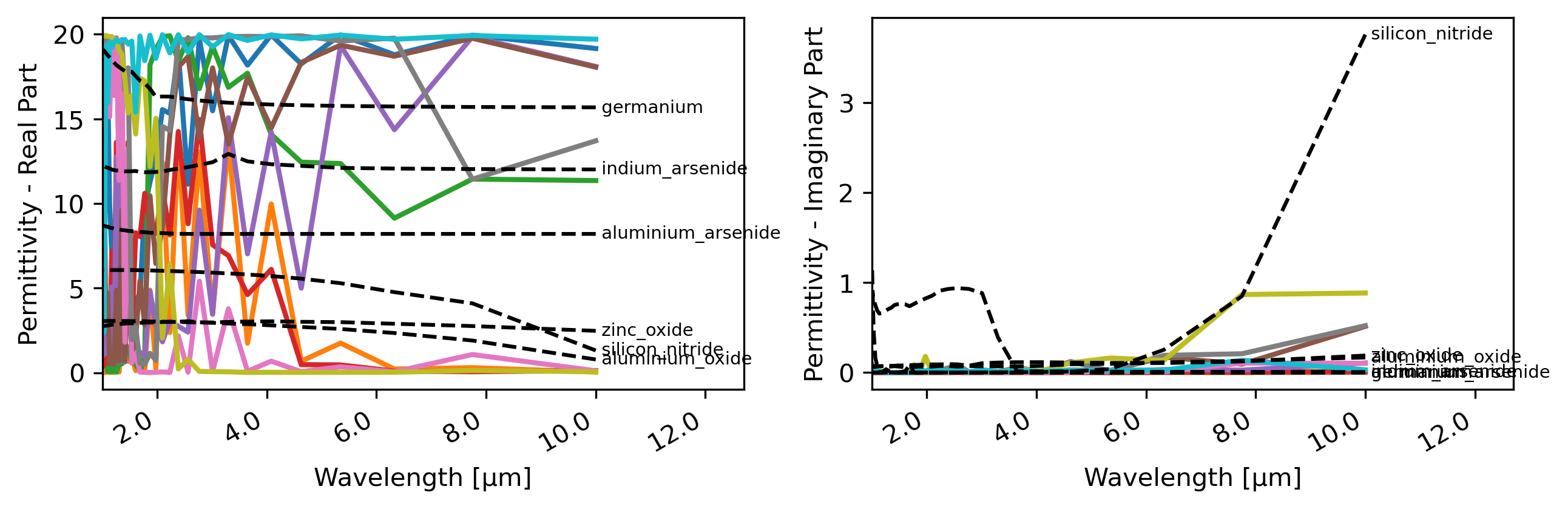}}
\caption{Inversion results for designing a bandpass filter with a ten layer stack using RCWA and regression; a) displays target and produced spectra, b) shows the utilized permittivity for each layer and closest material in NIDN (dashed)}
\label{fig:reg_filter}
\end{figure}

\subsection*{Filter Design}
The first test case closer to a potential application is the design of a bandpass filter. In this case, the target spectral properties are defined manually. Hence, the spectral characteristics may actually not be obtainable with the material properties the network is allowed to utilize. A stack of ten uniform layers is used to design the material. Due to the larger computational cost of FDTD, only RCWA was studied in this case. Also, the classification for this test case did not converge - likely because the available materials in database did not suffice to find a solution for this problem. 

\subsubsection*{Regression}
With the regression, the chosen range for the relative permittivity is $\epsilon_r \in [0 , 20 + 1i]$. The achieved spectral characteristics with regression and RCWA of the designed filter are shown in Figure \ref{fig:reg_filter}. The designed material captures the gap at the chosen wavelength around 1550 nm very well. Notably, the utilized permittivity is not similar to those of existing materials with erratic peaks in the real component of the permittivity for some of the layers.

\subsection*{Perfect Anti-Reflection Coating}
This second test case related to an application explores the scenario of building a perfect anti-reflection on top of a substrate with relative permittivity $\epsilon_r = 16$. This is closely related to the research described by Kim \& Park \cite{kim2013perfect}. Thus, we use a stack of eight \SI{0.1}{\micro\metre} thick layers on top of the higher permittivity substrate to achieve an anti-reflection coating (ARC). \\

\begin{figure}[!htbp]
\centering
\subcaptionbox{Target and Obtained Spectrum}
{\includegraphics[width=\customPlotWidth\linewidth]{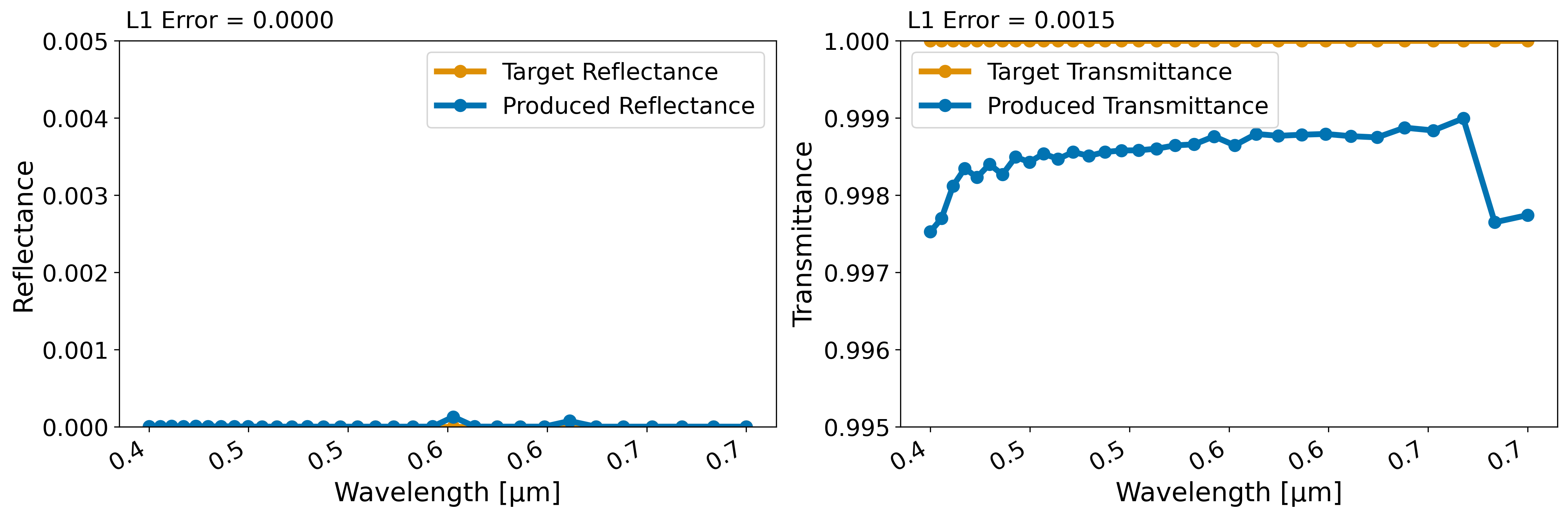}}
\subcaptionbox{Predicted Permittivity}
{\includegraphics[width=\customPlotWidth\linewidth]{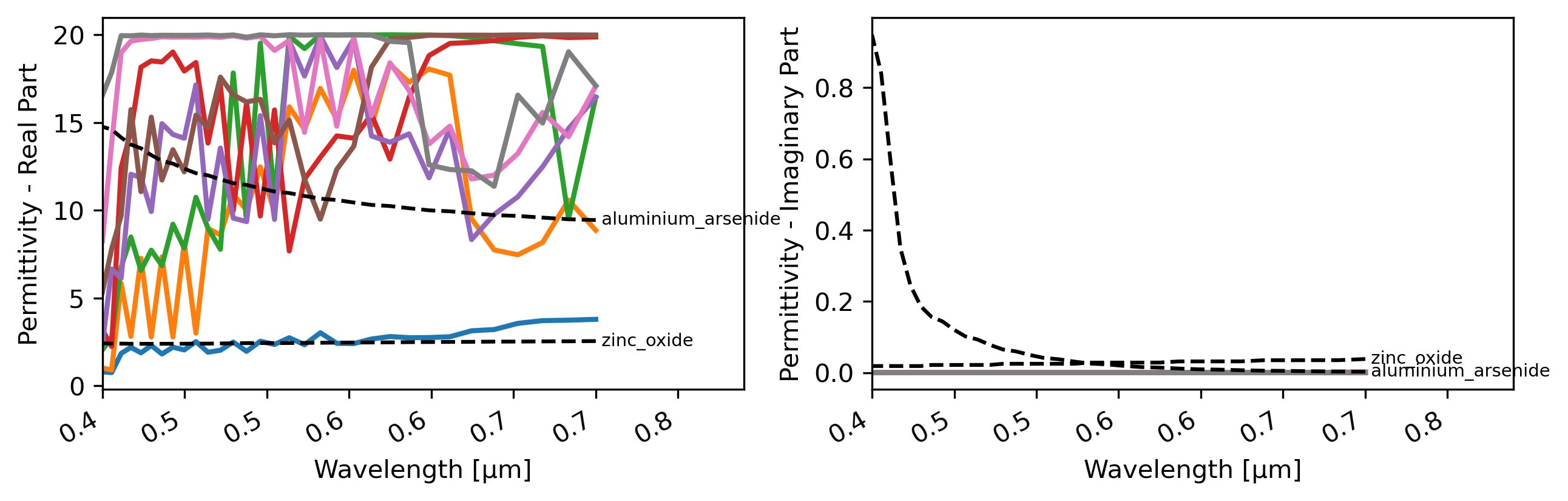}}
\caption{Inversion results for designing an anti-reflection coating with an eight layer stack using RCWA and regression; a) displays target and produced spectra, b) shows the utilized permittivity for each layer and closest material in NIDN (dashed)}
\label{fig:reg_AR}
\end{figure}

\subsubsection*{Regression}
For the regression with RCWA, Figure \ref{fig:reg_AR} displays the achieved spectral characteristics. The obtained transmittance is above 0.9975 for the entire range between \SIlist{400;700}{\nano\metre}, a peak transmittance of 0.9990 is achieved, the average is 0.9985. However, once again we can observe -- as seen in Figure \ref{fig:reg_AR}b  -- that the chosen permittivity does not resemble real material properties. 

\begin{figure}[!htbp]
\centering
\subcaptionbox{Target and Obtained Spectrum}
{\includegraphics[width=\customPlotWidth\linewidth]{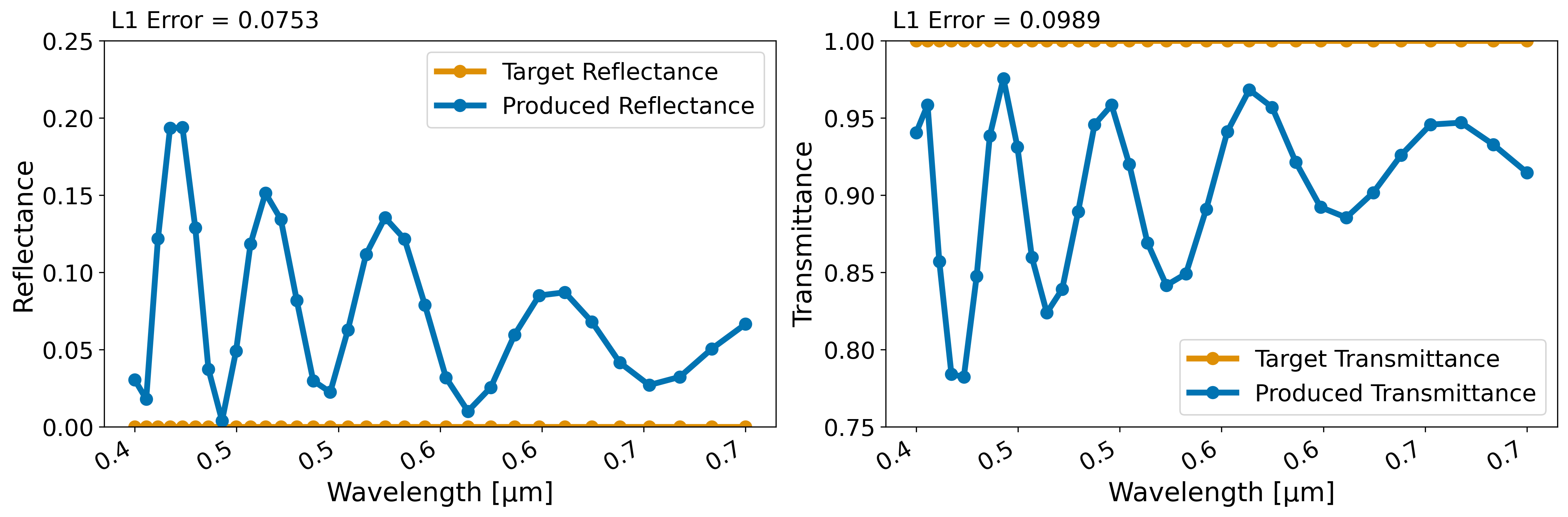}}
\subcaptionbox{Predicted Permittivity}
{\includegraphics[width=\customPlotWidth\linewidth]{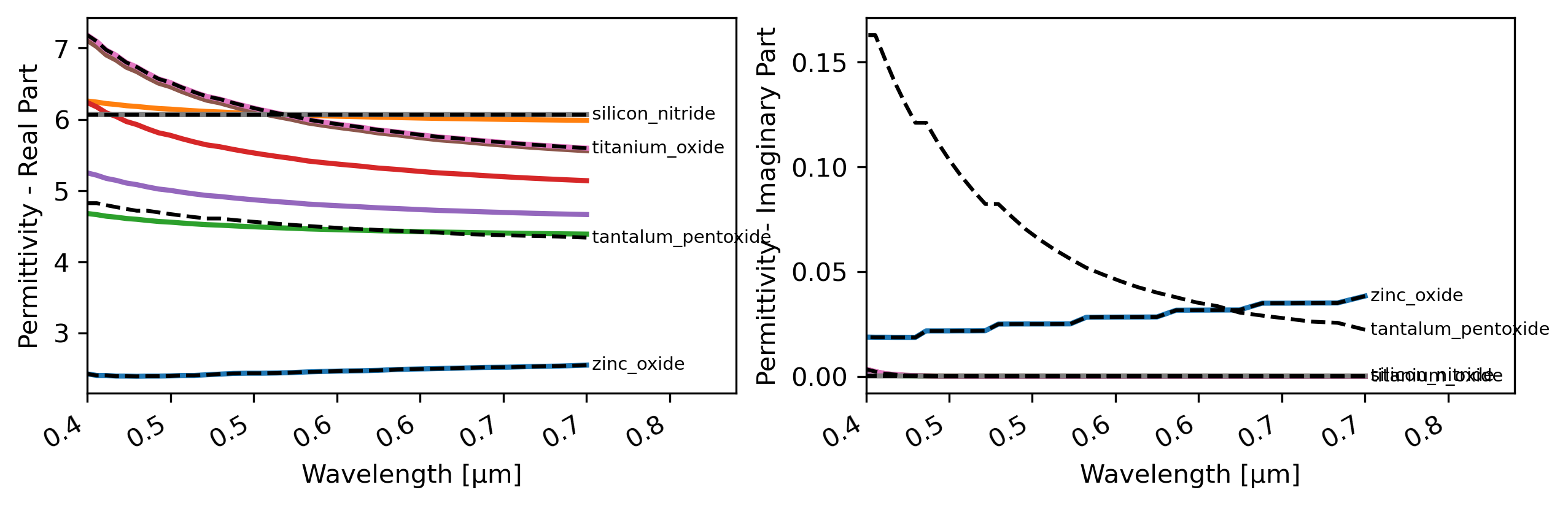}}
\caption{Inversion results for designing an anti-reflection coating with an eight layer stack using RCWA and classification; a) displays target and produced spectra, b) shows the utilized permittivity for each layer and closest material in NIDN (dashed)}
\label{fig:class_AR}
\end{figure}

\subsubsection*{Classification}
With the classification approach, the material design is still successful as can be seen in Figure \ref{fig:class_AR}. On average, the transmittance is 0.9011 with all values being over 0.7822 and a peak at 0.9756. As discernible in Figure \ref{fig:class_AR}b, the chosen materials for the layers in this approach are similar to existing ones such as Silicon Nitride and \tio{}. 

\section*{Discussion}
Overall, the results presented in this work clearly highlight the capabilities and potential of NIDN. We have shown it is able to correctly reproduce the material properties of several synthetic examples, which demonstrates that the inversion and training is successful for both, RCWA and FDTD. Similarly, the creation of both a \SI{1550}{\nano\metre} filter and a perfect ARC are achieved. However, especially the regression approach is susceptible to finding solutions which are not plausible for real materials. The main reason for this is the lack of constraints on the chosen permittivity. Currently, the constraints on physical plausibility in NIDN are implicitly given through the numerical simulations for determining the spectral characteristics. An additional component that checks if the predicted permittivity matches a realistic permittivity may be needed. The classification approach is able to partially remedy this, but for the real-world examples of filter design and ARC it is less effective at matching desired spectral characteristics. This is likely due to the limited number (12) of materials currently available in NIDN. The number can be increased, however, this also creates a harder classification problem which may in turn require finding an optimal trade-off of the number and diversity of available materials vs. the design limitations induced by fewer available materials. \par
One aspect where NIDN is more versatile than previous approaches \cite{Nadell_Meta_Trans_CST_FFDS_2019} is that no imposition of shape or form is required or performed. In fact, even the discretization into a grid is not a constraint as it is only required by the solver (RCWA or FDTD); the neural network output is spatially continuous. One thing that has been observed is, however, that the simpler case of having uniform layers benefits training convergence when utilizing RCWA. The reason for this is likely the need to iteratively solve an eigenvalue problem if non-uniform layers are present which may deteriorate gradients. This may be remedied, e.g., by returning to a stochastic setting where gradients are averaged over several tested materials to make the training more robust. This does come at an increased computational cost but may be parallelized. \par
In summary, NIDN demonstrates the flexibility of physics-based deep learning methods. The range of applications for the software goes well beyond the shown test cases of filters and ARC as it conceivably can be used for a broad range applications such as radiative materials \cite{Sun_Meta_OSR_FEM-FDTD_2017,Sun_Meta_OSR_FEM-FDTD_2018} or solar sails \cite{Jin_Meta_Sail_RCWA_AutoGrad_2020}. With RCWA and FDTD, two complementary solvers are integrated, RCWA being fast and well-suited for higher wavelength and FDTD being more potent and capable to investigate nanostructures in the visible and infrared spectrum. In the future, optimization and parallelization of the solvers will be critical to improve performance as especially training with FDTD for complex materials becomes prohibitively computationally expensive. Further, FDTD can also be used for the design of non-periodic structures which is a feature not yet implemented in NIDN. 

\section*{Methods}

\subsection*{Rigorous Coupled-Wave Analysis (RCWA)}
The RCWA method is a well-established tool for modeling scattering of stacked structures \cite{moharam1981rigorous}. In the NIDN implementation, single material as well as patterned layers can be stacked. A detailed overview of the capabilities can be found in the documentation of the Python module \textit{grcwa} \cite{Jin_Meta_Sail_RCWA_AutoGrad_2020} which NIDN's implementation is derived from. Note that the numerical stability of RCWA is dependent on the ratio of the size of the material grid to the investigated wavelength, i.e., for ratios above one RCWA may produce non-physical results. \\
The implementation in NIDN supports specifying a top or bottom layer in addition to the material; by default a vacuum layer is placed at both. The grid dimensions, number of layers and thicknesses of all layers can be specified manually. The default truncation order for simulations is 11 to keep computational costs low but it is configurable. The implementation employs a relaxation by adding a fictitious material absorption loss \cite{Jin_Meta_Sail_RCWA_AutoGrad_2020,liang2013formulation}. For now, we only investigated inversion with perpendicular light incidence angles and polarized light, however, both are also configurable in the RCWA implementation.

\subsection*{Finite-Difference Time-Domain (FDTD)}

FDTD is a simulation method that models the electric and magnetic field using a regular grid as a spatial discretization with a finite difference method \cite{Schneider2010ufdtd}. The electric and magnetic fields at each grid point are calculated with numerical approximations of Maxwell's equations; more specifically the electric field at each point is calculated using the curl of the magnetic field surrounding the point, and the magnetic field is calculated using the curl of the electric field surrounding the point.

The grid size is set to one tenth of the smallest wavelength used in the simulations, based on recommendations from the fdtd module's documentation.\footnote{\url{https://fdtd.readthedocs.io/en/latest/} Accessed: 2022-05-02} The time step is set according to the  Courant-Friedrichs-Lewy Condition \cite{de2013courant}, to ensure numerical stability: 
\begin{equation}
    \Delta t = \frac{S_c\cdot \Delta x}{u}
\end{equation}
where $\Delta t$ is the time step, $u$ is the wave velocity, $\Delta x$ is the spatial resolution and $S_c = \sqrt{dim} = \sqrt{2}$ is the Courant number.

In this work, the FDTD simulations were setup with periodic boundaries in the directions orthogonal to the wave-propagation to simulate an infinite plane. In the direction of propagation, Perfectly Matching Layers (PML) were used. PMLs aim to absorb as much of the electromagnetic-wave as possible to avoid reflections at the boundaries of the grid. In the propagation direction, the grid is built up by a PML layer of 1.5 \textmu m, a vacuum layer of 1.0 \textmu m, the material(s) chosen with the specified thicknesses, a vacuum layer of 1.0 \textmu m and finally another PML layer of 1.5 \textmu m. The source is a line across the width of the grid, radiating a continuous wave, placed at the interface between the first PML-layer and the free-space layer.

We measure the electromagnetic wave at placed detectors in the FDTD simulations such that the signal is independent of the direction of the incoming wave. Due to this, a correction of the reflected signal is necessary as the raw signal contains both the forward-going wave before reflection and the reflected wave. For the correction, we use an identical simulation setup without a material, the so-called free-space case, and subtract the measured signal from this simulation from the one with a material.

The transmission and reflection coefficients are calculated by comparing the energy of the wave after the transmission and a free-space case where there is no loss from a material such that 
\begin{equation}
    T = \frac{P_i}{P_t} =  \frac{E_{0i}^2}{E_{0t}^2}
\end{equation}
where the subscripts $i$,$t$ and $0$ refer to incident wave, transmitted wave and amplitude of the wave, respectively. Since the mean squared value of an integer number of periods of a sine wave equals $\frac{E_0^2}{2}$, the mean squared value is used to calculate the transmission coefficients. The same is the case for the corrected reflection signal. No dispersion model has been implemented as frequencies are computed independently.
By running the simulation for all desired wavelengths, the coefficients from each simulation were accumulated to obtain a transmission and reflection spectrum. The absorption spectrum is calculated by $A = 1-T-R$.

The number of time steps were chosen such that both the reflected signal and transmitted signal have at least two peaks, such that the mean square can be calculated over at least one period. The loss from a material in FDTD comes from the conductivity parameter. This is included in the \textit{AbsorbingObject} from the fdtd module, which is the only object/material type currently implemented in NIDN. The conductivity $\sigma$ is set to $\sigma = \epsilon '' \cdot\epsilon_0\cdot \omega$

\subsection*{Model Inversion Approach}

The inversion process in NIDN is inspired by recent research related to physics-based deep learning \cite{thuerey2021physics} and in directly optimizing physical properties using differentiable simulations \cite{izzo2021geodesy}.  The main advantage of these approaches is that gradients in training neural networks are propagated directly through a numerical simulation (also referred to as forward model in the context of inverse problems). Thus, there is no need to build a dataset of any kind. In particular, the process in NIDN consists of the following steps which are also depicted in Figure \ref{fig:nidn}:
\begin{enumerate}
    \item Specify target spectral characteristics (reflectance, transmittance, absorptance) and material properties (size, allowed permittivity, simulation parameters)
    \item Randomly initialize a neural network that prompted with a position returns the material's permittivity at that position
    \item Perform an evaluation of the numerical model for all target frequencies, i.e., compute the spectral characteristics for the materials as described by the network
    \item Compute the loss, i.e., the difference in the target spectral characteristics and those obtained through the network's material
    \item Backpropagate the gradients and update the network parameters to iteratively minimize the difference between target and obtained spectral characteristics
    \item Until results are satisfactory, return to Step 3
\end{enumerate}
After this process, the network is directly encoding the designed material by providing a permittivity for each position in the material. Herein, the conversion from a position to a permittivity is a particularly critical aspect to allow successful inverse design which will now be described in more detail.
\subsection*{Direct Permittivity Inference}
NIDN implements two different ways to determine the relative permittivity $\epsilon_r^P$ at a point $P = (x,y,z,f)$ where $x$ and $y$ are the point's location in a layer, $z$ is the index of the layer it is on and $f$ is the frequency at which we are evaluating. The first way is referred to as regression because the network herein directly infers a permittivity mapping $P \mapsto \epsilon_r^P$. The values are then clipped at some specified range to restrict it to physically plausible permittivity values. However, as can be seen in the results section, the frequency-dependent permittivity is still fairly unconstrained, allowing for materials that are not plausible as they, e.g., do not necessarily satisfy the Kramers-Kronig relationship \cite{lucarini2005kramers}. But, the advantage of this approach is the direct optimization of the material permittivity to obtain a solution that -- at least with regard to the Maxwell solver -- is plausible and can give a first idea of the hypothetical feasibility of some spectral characteristic assuming few constraints on permittivity. Permittivity values close to 0 are clipped to $\pm 0.01$ to avoid singularities in the Maxwell solvers.
\subsection*{Material Classification}
In order to obtain materials of higher physical fidelity, NIDN implements a second way of encoding material permittivity in the neural network. Naively, one may assume that it would be easiest to have the network perform a classification by assigning one specific material (and hence its permittivity) to each grid cell. Unfortunately, this is infeasible as the necessary \textit{argmax} operation in a classification (to select the most probable material) is not differentiable and thus would break the gradient flow. As an alternative, NIDN features a collection of $N$ - at time of writing twelve - materials from which we determine the relative permittivity $\epsilon_r$ at a point $P$ through a linear combination such that
\begin{equation}
    \epsilon_r^P  = \sum_{i=1}^N y^P \epsilon_r^i,
\end{equation}
where $y^P \in [0,1]^N$ is the neural network output for point $P$ and its $i$-th entry describes the probability of the $i$-th material at the point $P$ and $\epsilon_r^i$ is the relative permittivity of the $i$-th material in NIDN, respectively. Thus, the produced permittivity is strongly resemblant of real experimental permittivity as can be seen in the results for the classification approach. Furthermore, to push the network towards picking one material without breaking gradient flow, the output $y^P$ is passed through a softmax function with $\beta=16$.
\subsection*{Neural Network Training}
The neural network training in NIDN is inspired by several previous works in the field of differentiable models \cite{izzo2021geodesy,von2021study,mildenhall2020nerf,sitzmann2020implicit}. Thus, dense neural networks, namely the NeRF \cite{mildenhall2020nerf} and Siren \cite{sitzmann2020implicit} architecture are employed. NIDN allows a variety of losses such as absolute mean error or L2 error. For the classification, an additional regularization loss is used penalizing probabilities for their distance from the values 0 and 1. Training uses the Adam optimizer,  a default learning rate of $8\cdot10^{-5}$ and a learning rate scheduler to reduce the learning rate upon encountering a plateauing loss. 
\subsection*{Software Structure}
In general, NIDN enables users with two potential use cases: The first is to run forward model simulations using either RCWA or FDTD to obtain the spectral characteristics of a material. The second is the inverse design using neural network models to describe a material structure that produces certain spectral characteristics. NIDN's software repository has \textit{Jupyter} notebooks describing both processes in detail for RCWA and FDTD to enable users to run their own experiments. The project is open to contributions and under a permissive GPL-3 open-source license to allow customization. \\
NIDN is designed with modularity in mind, allowing for arbitrary neural network models and Maxwell solvers to be plugged in. It uses a continuous integration methodology with automated unit tests to ensure the correctness of the implementation. Several plotting utilities, such as those used for the figures in this work are available.  It is the authors' hope that the described steps and features will allow others to build and improve upon it.
\FloatBarrier
\bibliography{sample}

\begin{thebibliography}{10}
\urlstyle{rm}
\expandafter\ifx\csname url\endcsname\relax
  \def\url#1{\texttt{#1}}\fi
\expandafter\ifx\csname urlprefix\endcsname\relax\def\urlprefix{URL }\fi
\expandafter\ifx\csname doiprefix\endcsname\relax\def\doiprefix{DOI: }\fi
\providecommand{\bibinfo}[2]{#2}
\providecommand{\eprint}[2][]{\url{#2}}

\bibitem{wei2019machine}
\bibinfo{author}{Wei, J.} \emph{et~al.}
\newblock \bibinfo{journal}{\bibinfo{title}{Machine learning in materials
  science}}.
\newblock {\emph{\JournalTitle{InfoMat}}} \textbf{\bibinfo{volume}{1}},
  \bibinfo{pages}{338--358} (\bibinfo{year}{2019}).

\bibitem{liu2017materials}
\bibinfo{author}{Liu, Y.}, \bibinfo{author}{Zhao, T.}, \bibinfo{author}{Ju, W.}
  \& \bibinfo{author}{Shi, S.}
\newblock \bibinfo{journal}{\bibinfo{title}{Materials discovery and design
  using machine learning}}.
\newblock {\emph{\JournalTitle{Journal of Materiomics}}}
  \textbf{\bibinfo{volume}{3}}, \bibinfo{pages}{159--177}
  (\bibinfo{year}{2017}).

\bibitem{jiang2021deep}
\bibinfo{author}{Jiang, J.}, \bibinfo{author}{Chen, M.} \&
  \bibinfo{author}{Fan, J.~A.}
\newblock \bibinfo{journal}{\bibinfo{title}{Deep neural networks for the
  evaluation and design of photonic devices}}.
\newblock {\emph{\JournalTitle{Nature Reviews Materials}}}
  \textbf{\bibinfo{volume}{6}}, \bibinfo{pages}{679--700}
  (\bibinfo{year}{2021}).

\bibitem{Liu_Meta_Trans_FEM_GAN_2018}
\bibinfo{author}{Liu, Z.}, \bibinfo{author}{Zhu, D.},
  \bibinfo{author}{Rodrigues, S.~P.}, \bibinfo{author}{Lee, K.-T.} \&
  \bibinfo{author}{Cai, W.}
\newblock \bibinfo{journal}{\bibinfo{title}{Generative model for the inverse
  design of metasurfaces}}.
\newblock {\emph{\JournalTitle{Nano Letters}}} \textbf{\bibinfo{volume}{18}},
  \bibinfo{pages}{6570--6576}, \doiprefix\url{10.1021/acs.nanolett.8b03171}
  (\bibinfo{year}{2018}).

\bibitem{Ma_Meta_Refl_FreqDomain_VAE_2019}
\bibinfo{author}{Ma, W.}, \bibinfo{author}{Cheng, F.}, \bibinfo{author}{Xu,
  Y.}, \bibinfo{author}{Wen, Q.} \& \bibinfo{author}{Liu, Y.}
\newblock \bibinfo{journal}{\bibinfo{title}{Probabilistic representation and
  inverse design of metamaterials based on a deep generative model with
  semi-supervised learning strategy}}.
\newblock {\emph{\JournalTitle{Advanced Materials}}}
  \textbf{\bibinfo{volume}{31}}, \bibinfo{pages}{1901111},
  \doiprefix\url{10.1002/adma.201901111} (\bibinfo{year}{2019}).

\bibitem{Tahersima_NanoPh_DM_FDTD_Backprop_2019}
\bibinfo{author}{Tahersima, M.~H.} \emph{et~al.}
\newblock \bibinfo{journal}{\bibinfo{title}{Deep neural network inverse design
  of integrated photonic power splitters}}.
\newblock {\emph{\JournalTitle{Scientific Reports}}}
  \textbf{\bibinfo{volume}{9}}, \doiprefix\url{10.1038/s41598-018-37952-2}
  (\bibinfo{year}{2019}).

\bibitem{Piggott_NanoPh_DM_FDFD_GradDescent_2017}
\bibinfo{author}{Piggott, A.~Y.}, \bibinfo{author}{Petykiewicz, J.},
  \bibinfo{author}{Su, L.} \& \bibinfo{author}{Vu{\v{c}}kovi{\'{c}}, J.}
\newblock \bibinfo{journal}{\bibinfo{title}{Fabrication-constrained
  nanophotonic inverse design}}.
\newblock {\emph{\JournalTitle{Scientific Reports}}}
  \textbf{\bibinfo{volume}{7}}, \doiprefix\url{10.1038/s41598-017-01939-2}
  (\bibinfo{year}{2017}).

\bibitem{Rodriguez_NanoPh_DM_FDFD_Backprop_2021}
\bibinfo{author}{Rodr{\'{\i}}guez, J.~A.} \emph{et~al.}
\newblock \bibinfo{journal}{\bibinfo{title}{Inverse design of plasma
  metamaterial devices for optical computing}}.
\newblock {\emph{\JournalTitle{Physical Review Applied}}}
  \textbf{\bibinfo{volume}{16}},
  \doiprefix\url{10.1103/physrevapplied.16.014023} (\bibinfo{year}{2021}).

\bibitem{sajedian2018finding}
\bibinfo{author}{Sajedian, I.}, \bibinfo{author}{Badloe, T.} \&
  \bibinfo{author}{Rho, J.}
\newblock \bibinfo{journal}{\bibinfo{title}{Finding the best design parameters
  for optical nanostructures using reinforcement learning}}.
\newblock {\emph{\JournalTitle{arXiv preprint arXiv:1810.10964}}}
  (\bibinfo{year}{2018}).

\bibitem{sajedian2019optimisation}
\bibinfo{author}{Sajedian, I.}, \bibinfo{author}{Badloe, T.} \&
  \bibinfo{author}{Rho, J.}
\newblock \bibinfo{journal}{\bibinfo{title}{Optimisation of colour generation
  from dielectric nanostructures using reinforcement learning}}.
\newblock {\emph{\JournalTitle{Optics express}}} \textbf{\bibinfo{volume}{27}},
  \bibinfo{pages}{5874--5883} (\bibinfo{year}{2019}).

\bibitem{Molesky_2018_Inv_des_Nat_Phot}
\bibinfo{author}{Molesky, S.} \emph{et~al.}
\newblock \bibinfo{journal}{\bibinfo{title}{Inverse design in nanophotonics}}.
\newblock {\emph{\JournalTitle{Nature Photonics}}}
  \textbf{\bibinfo{volume}{12}}, \bibinfo{pages}{659--670},
  \doiprefix\url{10.1038/s41566-018-0246-9} (\bibinfo{year}{2018}).

\bibitem{MORI_IKAROS_Sail_2010}
\bibinfo{author}{MORI, O.} \emph{et~al.}
\newblock \bibinfo{journal}{\bibinfo{title}{First solar power sail
  demonstration by {IKAROS}}}.
\newblock {\emph{\JournalTitle{Transactions of the Japan Society for
  Aeronautical and Space Sciences, Aerospace Technology Japan}}}
  \textbf{\bibinfo{volume}{8}}, \bibinfo{pages}{To{\_}4{\_}25--To{\_}4{\_}31},
  \doiprefix\url{10.2322/tastj.8.to_4_25} (\bibinfo{year}{2010}).

\bibitem{Ullery_Meta_Sail_FDTD_2018}
\bibinfo{author}{Ullery, D.~C.} \emph{et~al.}
\newblock \bibinfo{journal}{\bibinfo{title}{Strong solar radiation forces from
  anomalously reflecting metasurfaces for solar sail attitude control}}.
\newblock {\emph{\JournalTitle{Scientific Reports}}}
  \textbf{\bibinfo{volume}{8}}, \doiprefix\url{10.1038/s41598-018-28133-2}
  (\bibinfo{year}{2018}).

\bibitem{Jin_Meta_Sail_RCWA_AutoGrad_2020}
\bibinfo{author}{Jin, W.}, \bibinfo{author}{Li, W.},
  \bibinfo{author}{Orenstein, M.} \& \bibinfo{author}{Fan, S.}
\newblock \bibinfo{journal}{\bibinfo{title}{Inverse design of lightweight
  broadband reflector for relativistic lightsail propulsion}}.
\newblock {\emph{\JournalTitle{{ACS} Photonics}}} \textbf{\bibinfo{volume}{7}},
  \bibinfo{pages}{2350--2355}, \doiprefix\url{10.1021/acsphotonics.0c00768}
  (\bibinfo{year}{2020}).

\bibitem{Sun_Meta_OSR_FEM-FDTD_2017}
\bibinfo{author}{Sun, K.} \emph{et~al.}
\newblock \bibinfo{journal}{\bibinfo{title}{Metasurface optical solar
  reflectors using {AZO} transparent conducting oxides for radiative cooling of
  spacecraft}}.
\newblock {\emph{\JournalTitle{{ACS} Photonics}}} \textbf{\bibinfo{volume}{5}},
  \bibinfo{pages}{495--501}, \doiprefix\url{10.1021/acsphotonics.7b00991}
  (\bibinfo{year}{2017}).

\bibitem{Sun_Meta_OSR_FEM-FDTD_2018}
\bibinfo{author}{Sun, K.} \emph{et~al.}
\newblock \bibinfo{journal}{\bibinfo{title}{{VO$_{2}$} thermochromic
  metamaterial-based smart optical solar reflector}}.
\newblock {\emph{\JournalTitle{{ACS} Photonics}}} \textbf{\bibinfo{volume}{5}},
  \bibinfo{pages}{2280--2286}, \doiprefix\url{10.1021/acsphotonics.8b00119}
  (\bibinfo{year}{2018}).

\bibitem{Hossain_Meta_OSR_FEM_2015}
\bibinfo{author}{Hossain, M.~M.}, \bibinfo{author}{Jia, B.} \&
  \bibinfo{author}{Gu, M.}
\newblock \bibinfo{journal}{\bibinfo{title}{A metamaterial emitter for highly
  efficient radiative cooling}}.
\newblock {\emph{\JournalTitle{Advanced Optical Materials}}}
  \textbf{\bibinfo{volume}{3}}, \bibinfo{pages}{1047--1051},
  \doiprefix\url{10.1002/adom.201500119} (\bibinfo{year}{2015}).

\bibitem{jiang2020metanet}
\bibinfo{author}{Jiang, J.} \emph{et~al.}
\newblock \bibinfo{journal}{\bibinfo{title}{Metanet: a new paradigm for data
  sharing in photonics research}}.
\newblock {\emph{\JournalTitle{Optics Express}}} \textbf{\bibinfo{volume}{28}},
  \bibinfo{pages}{13670--13681} (\bibinfo{year}{2020}).

\bibitem{sullivan2013electromagnetic}
\bibinfo{author}{Sullivan, D.~M.}
\newblock \emph{\bibinfo{title}{Electromagnetic simulation using the FDTD
  method}} (\bibinfo{publisher}{John Wiley \& Sons}, \bibinfo{year}{2013}).

\bibitem{moharam1981rigorous}
\bibinfo{author}{Moharam, M.} \& \bibinfo{author}{Gaylord, T.}
\newblock \bibinfo{journal}{\bibinfo{title}{Rigorous coupled-wave analysis of
  planar-grating diffraction}}.
\newblock {\emph{\JournalTitle{JOSA}}} \textbf{\bibinfo{volume}{71}},
  \bibinfo{pages}{811--818} (\bibinfo{year}{1981}).

\bibitem{han2014numerical}
\bibinfo{author}{Han, K.} \& \bibinfo{author}{Chang, C.-H.}
\newblock \bibinfo{journal}{\bibinfo{title}{Numerical modeling of
  sub-wavelength anti-reflective structures for solar module applications}}.
\newblock {\emph{\JournalTitle{Nanomaterials}}} \textbf{\bibinfo{volume}{4}},
  \bibinfo{pages}{87--128} (\bibinfo{year}{2014}).

\bibitem{Jiang_Meta_OSR_FDTD_DBS_2021}
\bibinfo{author}{Jiang, X.} \emph{et~al.}
\newblock \bibinfo{journal}{\bibinfo{title}{Implementation of radiative cooling
  with an inverse-designed selective emitter}}.
\newblock {\emph{\JournalTitle{Optics Communications}}}
  \textbf{\bibinfo{volume}{497}}, \bibinfo{pages}{127209},
  \doiprefix\url{10.1016/j.optcom.2021.127209} (\bibinfo{year}{2021}).

\bibitem{thuerey2021physics}
\bibinfo{author}{Thuerey, N.} \emph{et~al.}
\newblock \bibinfo{journal}{\bibinfo{title}{Physics-based deep learning}}.
\newblock {\emph{\JournalTitle{arXiv preprint arXiv:2109.05237}}}
  (\bibinfo{year}{2021}).

\bibitem{Ma_Meta_Refl__VAE_2020}
\bibinfo{author}{Ma, W.} \& \bibinfo{author}{Liu, Y.}
\newblock \bibinfo{journal}{\bibinfo{title}{A data-efficient self-supervised
  deep learning model for design and characterization of nanophotonic
  structures}}.
\newblock {\emph{\JournalTitle{Science China Physics, Mechanics \& Astronomy}}}
  \textbf{\bibinfo{volume}{63}}, \doiprefix\url{10.1007/s11433-020-1575-2}
  (\bibinfo{year}{2020}).

\bibitem{Nadell_Meta_Trans_CST_FFDS_2019}
\bibinfo{author}{Nadell, C.~C.}, \bibinfo{author}{Huang, B.},
  \bibinfo{author}{Malof, J.~M.} \& \bibinfo{author}{Padilla, W.~J.}
\newblock \bibinfo{journal}{\bibinfo{title}{Deep learning for accelerated
  all-dielectric metasurface design}}.
\newblock {\emph{\JournalTitle{Optics Express}}} \textbf{\bibinfo{volume}{27}},
  \bibinfo{pages}{27523}, \doiprefix\url{10.1364/oe.27.027523}
  (\bibinfo{year}{2019}).

\bibitem{Colburn_Meta_Refl_RCWA_Autodiff_2021}
\bibinfo{author}{Colburn, S.} \& \bibinfo{author}{Majumdar, A.}
\newblock \bibinfo{journal}{\bibinfo{title}{Inverse design and flexible
  parameterization of meta-optics using algorithmic differentiation}}.
\newblock {\emph{\JournalTitle{Communications Physics}}}
  \textbf{\bibinfo{volume}{4}}, \doiprefix\url{10.1038/s42005-021-00568-6}
  (\bibinfo{year}{2021}).

\bibitem{Wojcieszak2016TransmissionTiO2}
\bibinfo{author}{Wojcieszak, D.}, \bibinfo{author}{Kaczmarek, D.} \&
  \bibinfo{author}{Domaradzki, J.}
\newblock \bibinfo{journal}{\bibinfo{title}{Analysis of surface properties of
  semiconducting (ti,pd,eu)ox thin films}}.
\newblock {\emph{\JournalTitle{Opto-Electronics Review}}}
  \textbf{\bibinfo{volume}{24}}, \bibinfo{pages}{15--19},
  \doiprefix\url{doi:10.1515/oere-2016-0003} (\bibinfo{year}{2016}).

\bibitem{kim2013perfect}
\bibinfo{author}{Kim, K.-H.} \& \bibinfo{author}{Park, Q.-H.}
\newblock \bibinfo{journal}{\bibinfo{title}{Perfect anti-reflection from first
  principles}}.
\newblock {\emph{\JournalTitle{Scientific reports}}}
  \textbf{\bibinfo{volume}{3}}, \bibinfo{pages}{1--5} (\bibinfo{year}{2013}).

\bibitem{liang2013formulation}
\bibinfo{author}{Liang, X.} \& \bibinfo{author}{Johnson, S.~G.}
\newblock \bibinfo{journal}{\bibinfo{title}{Formulation for scalable
  optimization of microcavities via the frequency-averaged local density of
  states}}.
\newblock {\emph{\JournalTitle{Optics express}}} \textbf{\bibinfo{volume}{21}},
  \bibinfo{pages}{30812--30841} (\bibinfo{year}{2013}).

\bibitem{Schneider2010ufdtd}
\bibinfo{author}{Schneider, J.~B.}
\newblock \bibinfo{title}{Understanding the finite-difference time-domain
  method} (\bibinfo{year}{2010}).

\bibitem{de2013courant}
\bibinfo{author}{De~Moura, C.~A.} \& \bibinfo{author}{Kubrusly, C.~S.}
\newblock \bibinfo{journal}{\bibinfo{title}{The courant--friedrichs--lewy (cfl)
  condition}}.
\newblock {\emph{\JournalTitle{AMC}}} \textbf{\bibinfo{volume}{10}}
  (\bibinfo{year}{2013}).

\bibitem{izzo2021geodesy}
\bibinfo{author}{Izzo, D.} \& \bibinfo{author}{G{\'o}mez, P.}
\newblock \bibinfo{journal}{\bibinfo{title}{Geodesy of irregular small bodies
  via neural density fields: geodesynets}}.
\newblock {\emph{\JournalTitle{arXiv preprint arXiv:2105.13031}}}
  (\bibinfo{year}{2021}).

\bibitem{lucarini2005kramers}
\bibinfo{author}{Lucarini, V.}, \bibinfo{author}{Saarinen, J.~J.},
  \bibinfo{author}{Peiponen, K.-E.} \& \bibinfo{author}{Vartiainen, E.~M.}
\newblock \emph{\bibinfo{title}{Kramers-Kronig relations in optical materials
  research}}, vol. \bibinfo{volume}{110} (\bibinfo{publisher}{Springer Science
  \& Business Media}, \bibinfo{year}{2005}).

\bibitem{von2021study}
\bibinfo{author}{von Looz, M.}, \bibinfo{author}{Gomez, P.} \&
  \bibinfo{author}{Izzo, D.}
\newblock \bibinfo{journal}{\bibinfo{title}{Study of the asteroid bennu using
  geodesyanns and osiris-rex data}}.
\newblock {\emph{\JournalTitle{arXiv preprint arXiv:2109.14427}}}
  (\bibinfo{year}{2021}).

\bibitem{mildenhall2020nerf}
\bibinfo{author}{Mildenhall, B.} \emph{et~al.}
\newblock \bibinfo{title}{Nerf: Representing scenes as neural radiance fields
  for view synthesis}.
\newblock In \emph{\bibinfo{booktitle}{European conference on computer
  vision}}, \bibinfo{pages}{405--421} (\bibinfo{organization}{Springer},
  \bibinfo{year}{2020}).

\bibitem{sitzmann2020implicit}
\bibinfo{author}{Sitzmann, V.}, \bibinfo{author}{Martel, J.},
  \bibinfo{author}{Bergman, A.}, \bibinfo{author}{Lindell, D.} \&
  \bibinfo{author}{Wetzstein, G.}
\newblock \bibinfo{journal}{\bibinfo{title}{Implicit neural representations
  with periodic activation functions}}.
\newblock {\emph{\JournalTitle{Advances in Neural Information Processing
  Systems}}} \textbf{\bibinfo{volume}{33}}, \bibinfo{pages}{7462--7473}
  (\bibinfo{year}{2020}).

\end{thebibliography}


\section*{Author contributions statement}
P.G. and H.H.T. conceived of the presented idea, P.G., H.H.T and T.B.S. developed the computational framework and performed computations, P.G. and T.B.S. conducted the simulations, H.H.T, T.B.S., D.A.E. and J.M.L. contributed to the interpretation of results. J.M.L. provided supporting software and calculations. P.G. took the lead in writing the manuscript. All authors provided critical feedback and helped shape the research, analysis and manuscript.

\section*{Data availability statement}
All code and results presented in this paper are available open-source and open-access in the associated GitHub repository under \url{https://github.com/esa/NIDN}. \textit{Jupyter} notebooks in the repository contain all data shown in this article.
\section*{Additional information}

The  authors declare no competing interests.

\appendix
\counterwithin{figure}{section}

\section{Supplementary Information}

\begin{figure}[ht]
\centering
\subcaptionbox{Uniform Grid Test Case}
{\includegraphics[width=0.45\linewidth]{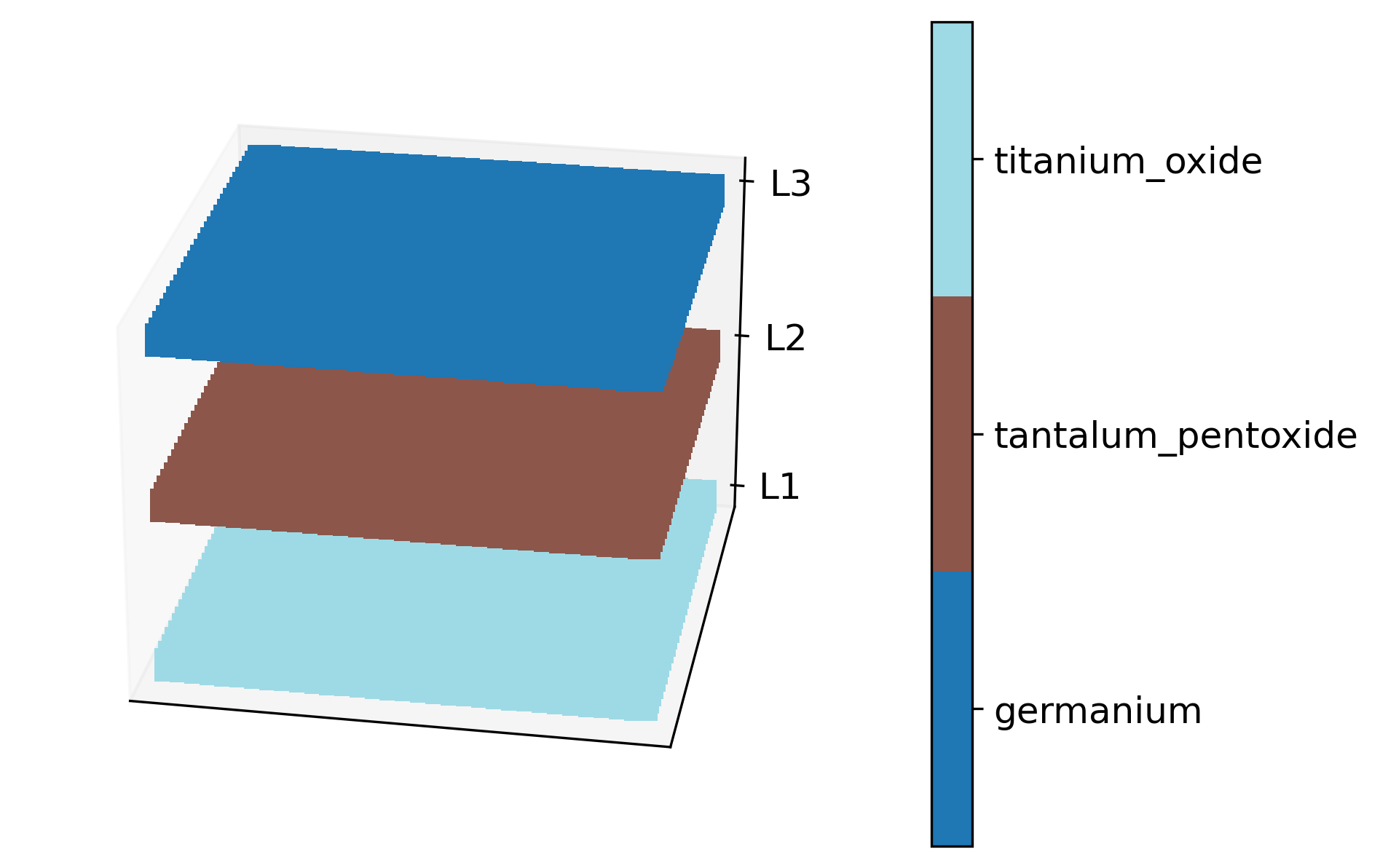}}
\subcaptionbox{Patterned Grid Test Case}
{\includegraphics[width=0.45\linewidth]{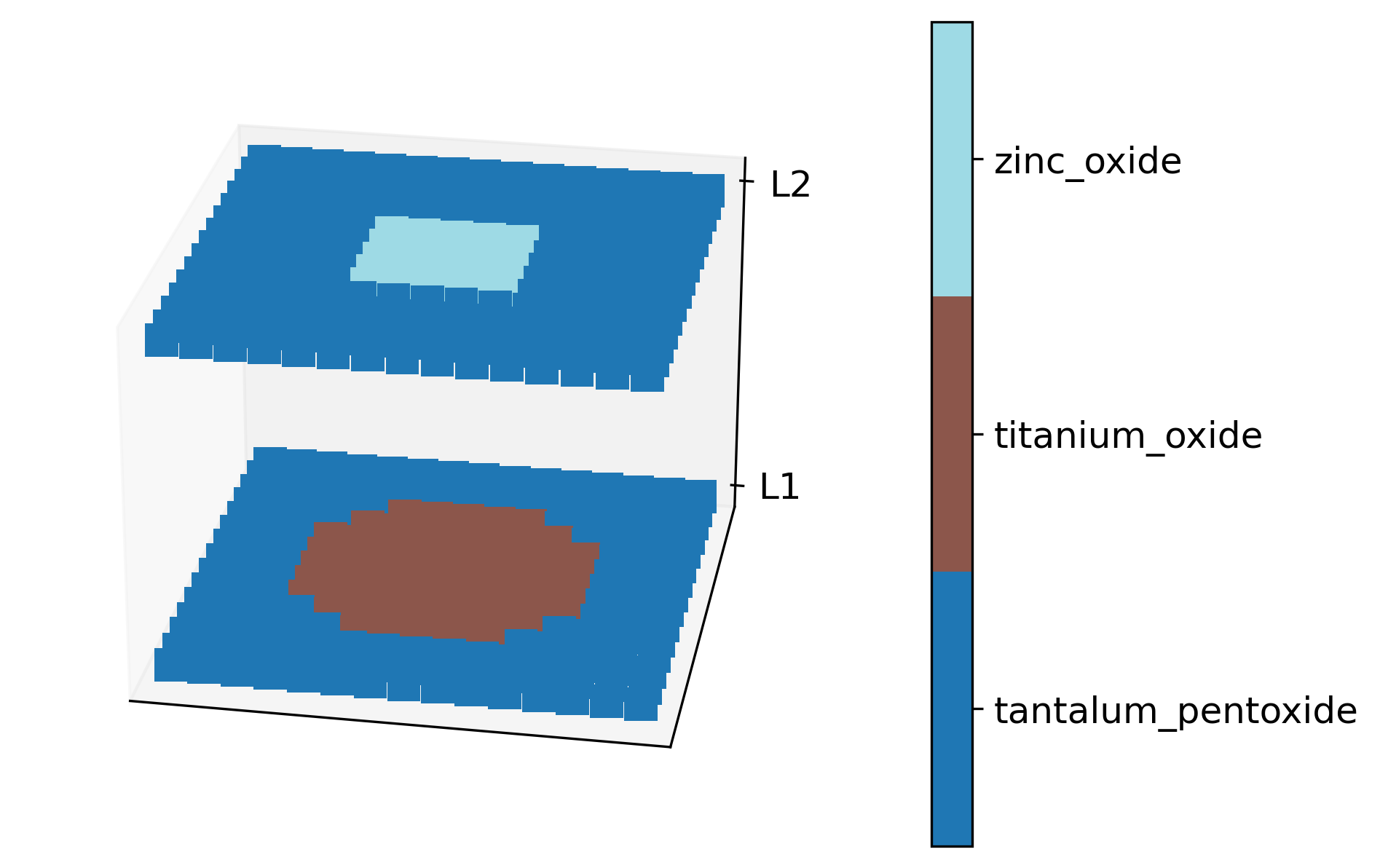}}
\caption{Geometry of test cases for validation against the original implementation of RCWA in the Python module grcwa. A three-layer homogeneous material and a two-layer patterned material are investigated.}
\label{fig:supp1}
\end{figure}

\begin{figure}[ht]
\centering
\includegraphics[width=\customPlotWidth\linewidth]{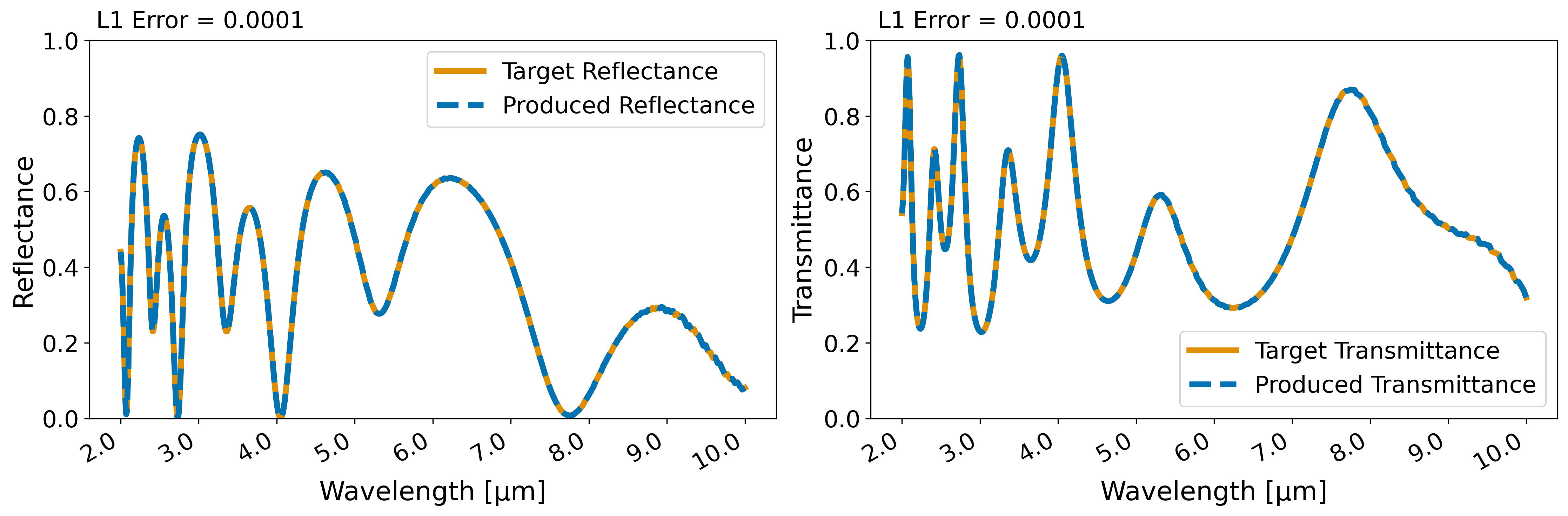}
\caption{A three-layer uniform material with a layer of Germanium, Titanium Dioxide (\tio) and Tantalum Pentoxide (\tape), respectively. The spectral characteristics are investigated at wavelengths between \SIlist{1;50}{\micro\metre}. The mean absolute error between GRCWA and NIDN is $7.90 \cdot 10^{-5}$. Overall, the spectra are very similar and differences are likely in the range of numerical errors.}
\label{fig:supp2}
\end{figure}

\begin{figure}[ht]
\centering
\includegraphics[width=\customPlotWidth\linewidth]{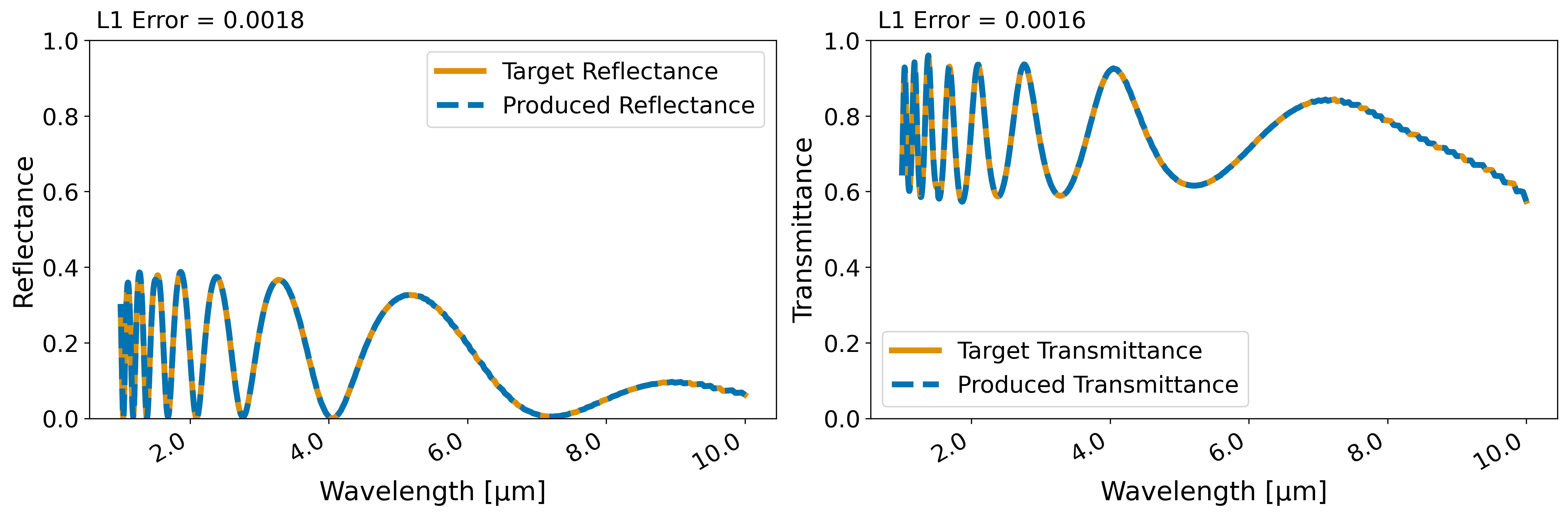}
\caption{The second test case is a material consisting of two patterned $\SI{0.1}{\micro\metre} \times \SI{0.1}{\micro\metre}$ layers consisting of \tape{} with an embedded square and circle component of ZnO and \tio{} in the center, respectively. A range of \SIrange{1}{10}{\micro\metre} is chosen. Obtained mean absolute error are between the spectra is $1.70 \cdot 10^{-3}$. Again, the obtained spectra match almost perfectly.
 \label{fig:grcwa_test_patterned}}
\end{figure}

We show the numerical comparison between the original implementation of RCWA in the Python package grcwa with the implementation in NIDN. Figure \ref{fig:supp1} contains the details of the uniform and patterned unit cells. Figure \ref{fig:supp2} shows the comparison in reflectance and transmittance of the unit cell from Figure \ref{fig:supp1}a and, correspondingly, Figure \ref{fig:grcwa_test_patterned} from Figure \ref{fig:supp1}b.

\begin{figure}[ht]
\centering
\subcaptionbox{fdtd vs. NIDN signal}
{\includegraphics[width=0.375\linewidth]{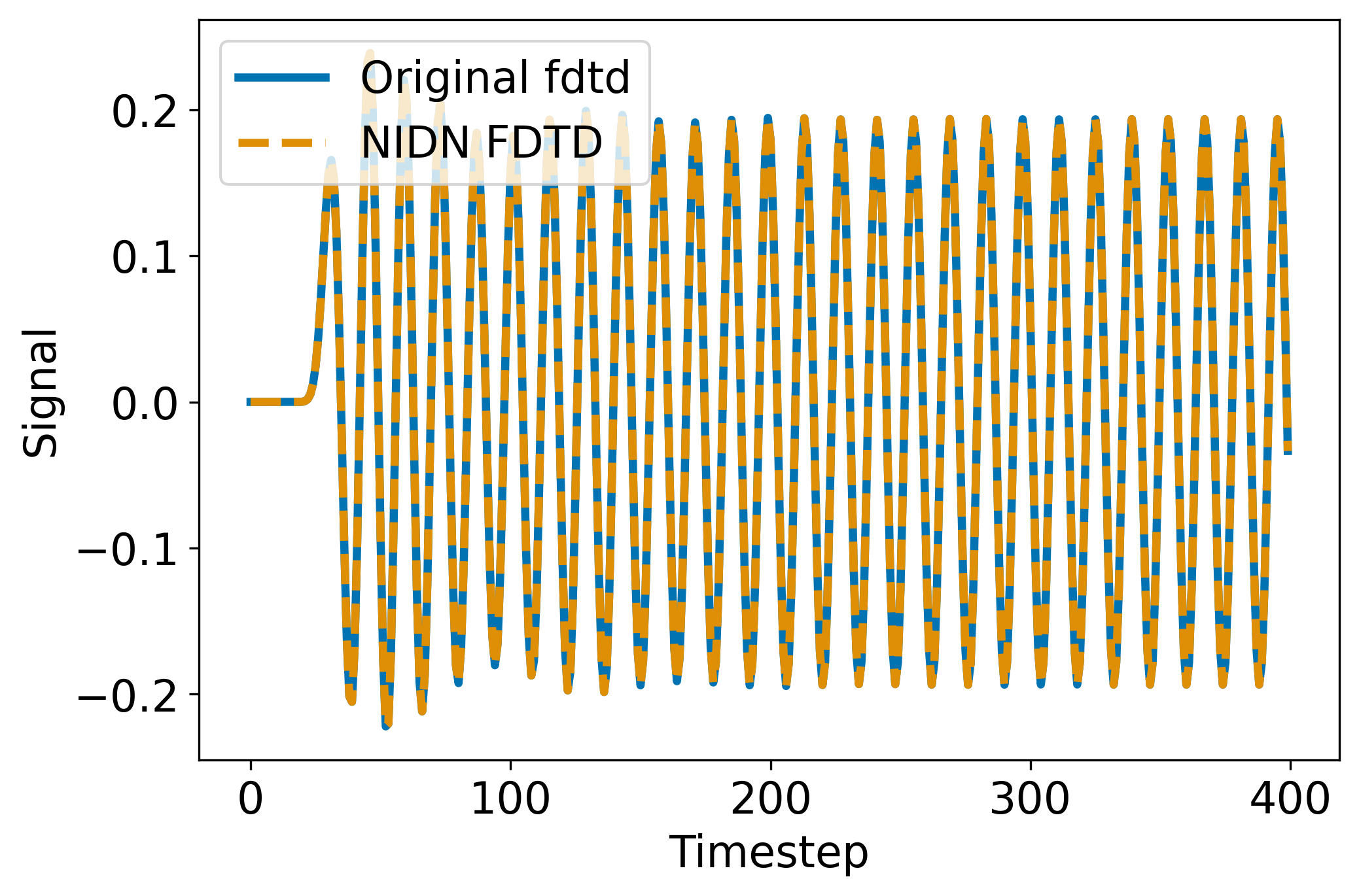}}
\subcaptionbox{Error between fdtd and NIDN}
{\includegraphics[width=0.375\linewidth]{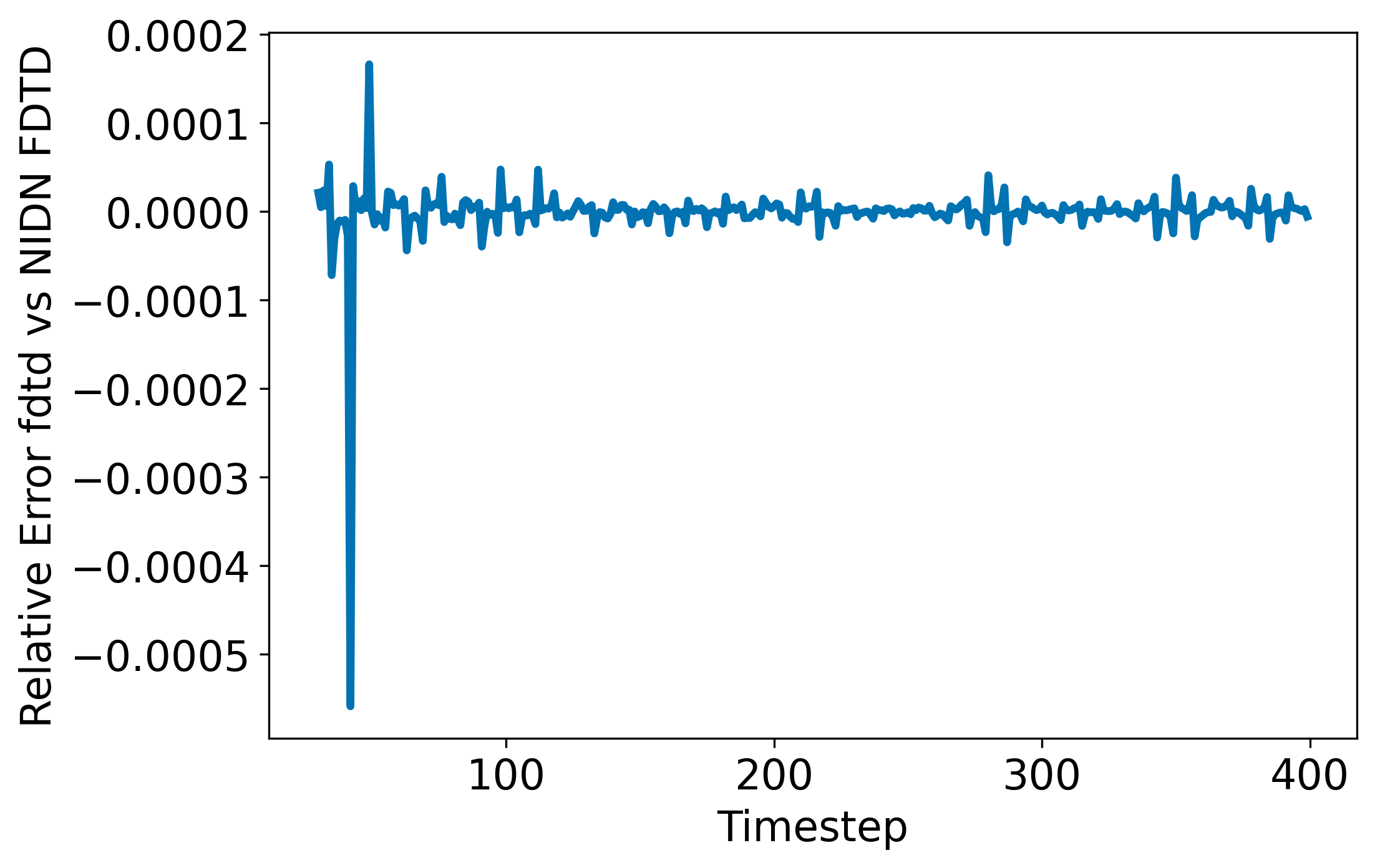}}
\caption{Direct comparison of a single \SI{300}{\nano\metre} thick \tio{} layer obtained with NIDN's FDTD implementation and the one from the fdtd Python module for a continuous light source at a wavelength \SI{1}{\micro\metre}. The mean absolute error between the original FDTD and NIDN is $6.88 \cdot 10^{-7}$. Overall, results remain virtually identical apart from numerical errors.}
\label{fig:fdtd2}
\end{figure}

The validation of the FDTD impelementation in NIDN with respect to the original package fdtd is shown in Figure \ref{fig:fdtd2} through the electric field in a TiO2 slab of 300 nm.


\end{document}